\documentclass[aps,prb,twocolumn,superscriptaddress,nofootinbib,longbibliography]{revtex4-2}
\usepackage[T1]{fontenc}
\usepackage[utf8]{inputenc}
\usepackage{array}
\usepackage{multirow}
\usepackage{graphicx}
\usepackage{amsmath,amsfonts,amssymb,color}
\usepackage{amsthm}
\usepackage{mathtools}
\usepackage{leftidx}
\usepackage{xcolor}
\usepackage{dcolumn}
\usepackage{bm}
\usepackage{epstopdf}
\usepackage{epsfig}
\usepackage{environ}
\usepackage{pdfcomment}
\usepackage{float}
\usepackage{setspace}
\usepackage{esint}
\usepackage{physics}
\usepackage{comment}
\hypersetup{colorlinks=true,citecolor=blue,
linkcolor=blue,urlcolor=blue,pdfstartview=FitH,
bookmarksopen=true}
\usepackage{cleveref}  

\begin{document}

\title{Kardar-Parisi-Zhang and glassy properties in 2D Anderson localization:\\ eigenstates and wave packets}

\author{Noam Izem}
\affiliation{Universit\'e Toulouse, CNRS, Laboratoire de Physique Th\'eorique, Toulouse, France}

\author{Bertrand Georgeot}
\affiliation{Universit\'e Toulouse, CNRS, Laboratoire de Physique Th\'eorique, Toulouse, France}

\author{Jiangbin Gong}
\affiliation{Centre for Quantum Technologies, National University of Singapore, Singapore 117543, Singapore}
\affiliation{Department of Physics, National University of Singapore, Singapore 117542, Singapore}
\affiliation{MajuLab, CNRS-UCA-SU-NUS-NTU International Joint Research Unit, Singapore}

\author{Gabriel Lemari\'e}\email{gabriel.lemarie@cnrs.fr}
\affiliation{MajuLab, CNRS-UCA-SU-NUS-NTU International Joint Research Unit, Singapore}
\affiliation{Centre for Quantum Technologies, National University of Singapore, Singapore 117543, Singapore}
\affiliation{Department of Physics, National University of Singapore, Singapore 117542, Singapore}
\affiliation{Universit\'e Toulouse, CNRS, Laboratoire de Physique Th\'eorique, Toulouse, France}
\affiliation{Universit\'e C\^ote d'Azur, CNRS, INPHYNI, France}
\date{\today}

\author{Sen Mu}\email{senmu@u.nus.edu}
\affiliation{Centre for Quantum Technologies, National University of Singapore, Singapore 117543, Singapore}
\affiliation{Max Planck Institute for the Physics of Complex Systems, N\"othnitzer Str.~38, 01187 Dresden, Germany}

\begin{abstract}
Despite decades of research, the universal nature of fluctuations in disordered quantum systems remains poorly understood. Here, we present extensive numerical evidence that fluctuations in two-dimensional (2D) Anderson localization belongs to the $(1+1)$-dimensional Kardar-Parisi-Zhang (KPZ) universality class. In turn, by adopting the KPZ framework, we gain fresh insight into the structure and phenomenology of Anderson localization itself.
We analyze both localized eigenstates and time-evolved wave packets, demonstrating that the fluctuation of their logarithmic density follows the KPZ scaling.
Moreover, we reveal that the internal structure of these eigenstates exhibits glassy features characteristic of the directed polymer problem, including the emergence of dominant paths together with pinning and avalanche behavior. Localization is not isotropic but organized along preferential branches of weaker confinement, corresponding to these dominant paths.
For localized wave packets, we further demonstrate that their spatial profiles obey a stretched-exponential form consistent with the KPZ scaling, while remaining fully compatible with the single-parameter scaling (SPS) hypothesis, a cornerstone of Anderson localization theory. Altogether, our results establish a unified KPZ framework for describing fluctuations and microscopic organization in 2D Anderson localization, revealing the glassy nature of localized states and providing new understanding into the universal structure of disordered quantum systems.
\end{abstract}

\maketitle

\section{Introduction}

Anderson localization is a fundamental phenomenon responsible for the absence of wave diffusion in disordered media \cite{anderson_absence_1958}. It plays a central role in understanding transport across a wide variety of systems, from solid-state materials \cite{dynes_localization_1984} to ultracold atomic gases \cite{Raizen_prl_1994, Raizen_prl_1995, billy_direct_2008,chabe_experimental_2008}, as well as for classical waves \cite{lagendijk_fifty_2009}.

Despite more than six decades of intensive research \cite{evers_anderson_2008,abrahams_50_2010}, a complete understanding of its properties in the strong-disorder regime and in dimensions higher than two remains elusive. In one-dimensional (1D) systems, Anderson localization is well characterized through supersymmetric field theory \cite{efetov_kinetics_1983,mirlin_statistical_1994,zirnbauer_super_1992,efetov_supersymmetry_1995,mirlin_statistics_2000} and the Dorokhov-Mello-Pereyra-Kumar (DMPK) framework \cite{dorokhov_transmission_1982,mello_macroscopic_1988,beenakker_random-matrix_1997}. These complementary approaches yield exact analytical results, most notably a log-normal distribution of wavefunction amplitudes, reflecting the intrinsically non-self-averaging character of localization \cite{mirlin_statistics_2000}.

The situation changes drastically in higher dimensions. Although it is well established that all single-particle eigenstates are localized for arbitrarily weak disorder in two-dimensional (2D) systems belonging to the orthogonal symmetry class \cite{pichard1981finite, mackinnon1981one}, a rigorous analytical treatment comparable to that available in 1D is still lacking. In particular, the nonlinear sigma model provides no closed analytical description of the strongly localized regime. Several theoretical and numerical efforts have sought to generalize the DMPK approach to higher dimensions \cite{muttalib_generalized_1999,muttalib_generalization_2002,douglas_generalized_2014,suslov_general_2018}, but an analytical characterization of the fluctuations in the 2D localized regime remains out of reach.

Early works on hopping transport \cite{nguyen_aaronov-bohm_1985,medina_quantum_1992,pietracaprina_forward_2016} and recent numerical studies \cite{prior_conductance_2005,somoza_universal_2007,lemarie_glassy_2019,mu_kardar-parisi-zhang_2024,Mu_prb_2025,swain_2d_2025} have revealed a deep and unexpected connection between 2D Anderson localization and the Kardar-Parisi-Zhang (KPZ) universality class in (1+1) dimensions. Originally introduced to describe the stochastic growth of rough interfaces \cite{kardar_dynamic_1986, takeuchi_appetizer_2018}, the KPZ framework has since become a cornerstone of non-equilibrium statistical physics, capturing universal fluctuation properties across a broad range of systems, both classical and quantum.

Fluctuations in the KPZ universality class are characterized by universal scaling exponents and non-Gaussian distributions, e.g.~Tracy-Widom distribution, well known as the distribution of the largest eigenvalue from the random-matrix theory \cite{kardar_statistical_2007,prahofer_statistical_2000,prahofer_universal_2000}. KPZ statistics has been experimentally observed in a variety of classical systems, including liquid-crystal interfaces and directed polymers in random media \cite{takeuchi_universal_2010,huergo_morphology_2010,myllys_scaling_2000,halpin-healy_kinetic_1995}.

Remarkably, KPZ scalings have also emerged in several quantum systems, such as one-dimensional quantum magnets \cite{ljubotina_kardar-parisi-zhang_2019,wei_quantum_2022}, random unitary circuits \cite{nahum_operator_2018}, and driven-dissipative quantum fluids \cite{comaron_dynamical_2018,fontaine_kardarparisizhang_2022,deligiannis_kardar-parisi-zhang_2022,amelio_kardar-parisi-zhang_2024}.
In the context of Anderson localization, pioneering numerical studies demonstrated that the fluctuations of the logarithm of the conductance in 2D disordered systems follow the KPZ scaling \cite{prior_conductance_2005,somoza_universal_2007}; see also Refs.~\cite{lemarie_glassy_2019,swain_2d_2025}.

In a recent letter, some of the authors of this work demonstrated that the logarithm of the density of 2D localized wave packets, relevant for experiments with e.g. ultrasound or atomic matter waves, also obeys the KPZ statistics \cite{mu_kardar-parisi-zhang_2024}.
In the present extended work, we provide a more comprehensive analysis of this connection. In particular, we employ a numerical scheme that enables high precision computations for the unitary time evolution under Hamiltonian. This level of control is essential for resolving the KPZ fluctuations that arise in the exponentially small tails of localized wave packets. Our results provide compelling evidence that the single-parameter scaling (SPS) hypothesis \cite{abrahams_scaling_1979, evers_anderson_2008, pichard1981finite, mackinnon1981one}, a cornerstone of Anderson localization theory, is fully consistent with the emergence of KPZ fluctuations in two dimensions.

Importantly, we extend the analogy with KPZ physics beyond transport observables to the most fundamental objects in Anderson localization: the eigenstates themselves.
A central question we address concerns how an individual eigenstate localizes within a given disorder configuration. This problem has recently been explored through the localization-landscape approach, which describes the spatial position of the localization center by constructing an effective confining potential \cite{filoche2012universal, arnold2016effective, PhysRevB.109.L220202, herviou20202, kakoi2023stochastic, skipetrov2024higher}. Here, we address a complementary aspect of this question by focusing on the spatial organization far from the localization center.

In the strongly localized regime where KPZ fluctuations arise, approximate arguments suggest an analogy with the directed polymer (DP) problem \cite{nguyen_aaronov-bohm_1985,medina_quantum_1992,pietracaprina_forward_2016}. The DP problem is a classical statistical model that not only belongs to the KPZ universality class but also exhibits the glassy properties of pinned elastic manifolds, such as pinning and avalanche phenomena \cite{mezard_glassy_1990, fisher1991directed, wiese2022theory}. In its glassy phase, the polymer becomes trapped along an optimal path that is stable against small perturbations (pinning) but undergoes abrupt rearrangements, a.k.a. avalanches, when external parameters are varied continuously. Similar glassy behavior of dominant paths has recently been characterized in quantum conductance studies of Anderson insulators \cite{lemarie_glassy_2019}.

This glassy perspective provides a powerful framework to elucidate the microscopic structure of localized eigenstates at strong disorder, complementing landscape-based approaches that describe localization centers. We show that localized eigenstates in 2D indeed exhibit glassy features characteristic of the DP analogy, with the emergence of dominant paths displaying pinning and avalanche behavior. Localization is not isotropic but organized along preferential branches of weaker localization, which correspond to the dominant paths selected by the underlying glassy physics of the directed-polymer problem (see also \cite{PhysRevResearch.2.012020, PhysRevB.106.214202, derrida1988polymers} for an even stronger form of anisotropy on random graphs).

Altogether, these findings deepen our understanding of both localized eigenstates and wave packets in 2D Anderson localization under strong disorder, and highlight the unifying role of KPZ and DP universality in unveiling the microscopic structure and fluctuation properties of Anderson localization.

Our paper is organized as follows. In Section~\ref{model} we describe the 2D Anderson model and the numerical strategy used to compute the eigenstates and wave packet dynamics. In Section~\ref{sec:Eigenstates} we present the properties of localized eigenstates and demonstrates that their logarithmic density fluctuations exhibit the KPZ scaling. In Section~\ref{sec:dominant_paths} we reveal the microscopic structure of the localized eigenstates by identifying dominant paths and characterizing their glassy properties, namely pinning and avalanches. In Section~\ref{Analytical_form} we relate the eigenstate correlations to long-time evolved wave packets and demonstrate that the former display the KPZ scaling as well. We then review the expressions proposed for the localized wave packets and present numerical results across a broad range of disorder strengths. In particular, the stretched exponential form of the localized wave packets motivated by KPZ physics is found compatible with the celebrated single-parameter scaling hypothesis of Anderson localization. Finally, in Section~\ref{conclusion} we summarize our findings and outline promising directions for future research.

\section{Model and numerical strategy}
\label{model}

We consider the standard tight-binding model for Anderson localization on a square lattice of size $N=L\times L$, described by the Hamiltonian
\begin{equation}
H = -\sum_{\langle \boldsymbol r, \boldsymbol r^\prime \rangle} 
\left( c^\dagger_{\boldsymbol r} c_{\boldsymbol r^\prime} 
+ c^\dagger_{\boldsymbol r^\prime} c_{\boldsymbol r} \right) 
+ \sum_{\boldsymbol r} w_{\boldsymbol r} n_{\boldsymbol r},
\label{H_AL}
\end{equation}
where $c^\dagger_{\boldsymbol r}$ and $c_{\boldsymbol r}$ are creation and annihilation operators, and 
$n_{\boldsymbol r} = c^\dagger_{\boldsymbol r} c_{\boldsymbol r}$ is the number operator at site 
$\boldsymbol r = (x, y)$.  
The sum $\langle \boldsymbol{r}, \boldsymbol{r}^\prime \rangle$ runs over nearest-neighbor pairs.
Onsite potentials $w_{\boldsymbol r}$ are identically and independently drawn from a uniform distribution 
in $[-W/2, W/2]$, where $W > 0$ defines the disorder strength.  

\subsection{Eigenstate computation}
\label{eigenstate_computation}

The first goal of this study is to characterize spatial fluctuations of eigenstates defined by:
\begin{equation}
H |\Psi_n\rangle = E_n |\Psi_n\rangle,
\end{equation}
where $|\Psi_n\rangle$ are eigenstates with corresponding eigenenergies $E_n$.  We diagonalize the Hamiltonian in Eq.~\eqref{H_AL} using two complementary methods. For systems up to $N = 64 \times 64$ sites, we perform \textit{exact diagonalization} using LAPACK routines, obtaining the full spectrum and eigenstates to machine precision.  
For larger lattices, we employ \textit{sparse diagonalization} via \texttt{scipy.sparse.linalg.eigsh}, which implements the implicitly restarted Arnoldi algorithm \cite{lehoucq_arpack_1998,noauthor_opencollabarpack-ng_2025}.  
This Krylov subspace projection method efficiently computes a subset of eigenvalues and eigenvectors near a chosen target energy without forming the full matrix explicitly.  
Using this approach, we reach system sizes up to $512 \times 512$, with most simulations at $256 \times 256$, balancing computational cost and spatial resolution.  
The number of eigenstates retained depends on the specific observable and required energy resolution.  

We note two limitations in computing the eigenstates. First, finite system size imposes a lower bound on disorder strength $W$: the localization length $\xi$ must remain much smaller than $L$ for eigenstates to exhibit exponential localization. Second, round-off errors in double-precision arithmetic limit the smallest resolvable amplitudes, values below approximately $10^{-16}$ become indistinguishable from numerical noise.  
Since eigenstate amplitudes decay exponentially, this sets an upper bound on $W$.  
While higher-precision arithmetic could extend the accessible range, the associated computational cost becomes prohibitive for large 2D systems.  
Within these constraints, the dense and sparse diagonalization methods provide accurate eigenstates, enabling a detailed study of fluctuation scaling and spatial structure in 2D Anderson localization.  

\subsection{Wave packet dynamics}

The second goal of this work is to investigate spatial fluctuations in long-time evolved, localized wave packets.  
The time evolution of the wave function obeys:
\begin{equation}
|\psi^t\rangle = e^{-i H t} |\psi^0\rangle,
\label{eq: se}
\end{equation}
where we set $\hbar = 1$.  

\subsubsection{Initial conditions}

We consider two distinct initial conditions for $|\psi^0\rangle$:
\begin{eqnarray}
\text{point:} &\quad& \psi^0(\boldsymbol{r}) = \delta(\boldsymbol{r}),\quad r = \sqrt{x^2+y^2}, 
\label{Eq:inicirc} \\
\text{line:} &\quad& \psi^0(x, k_y) = \frac{1}{\sqrt{L}} \delta(x)\,\delta(k_y),\quad r = |x|.
\label{Eq:iniflat}
\end{eqnarray}
Here, $r$ denotes the Euclidean distance along the direction of localization, over which the wave amplitude decays exponentially.  
The \textit{point} condition corresponds to a spatial delta function, while the \textit{line} condition describes a state localized along $x$ and extended along $y$.  
These configurations correspond, respectively, to circular and flat initial conditions in the mapped interface growth processes, as established in the previous work~\cite{mu_kardar-parisi-zhang_2024,takeuchi_evidence_2012}.  

\subsection{Time evolution via the scaling-and-squaring algorithm} \label{sec:timeevol}

We employ the \texttt{scipy.sparse.linalg.expm\_multiply} routine from the SciPy library~\cite{higham_computing_2010,al-mohy_computing_2011} to compute the time-evolved wave function governed by Eq.~\eqref{eq: se}.  
This method combines the scaling-and-squaring technique with a truncated Taylor expansion, making it well suited for large sparse Hamiltonians such as the model considered here.

The algorithm rescales the matrix $A = -iHt$ by a factor $2^{-s}$, with $s$ chosen so that the norm of the scaled matrix is small enough for an accurate truncated Taylor approximation of $e^{A/2^s}$.  
The resulting matrix is then squared $s$ times to reconstruct $e^A$.  
The expansion
\begin{equation}
e^{A} = I + A + \frac{A^2}{2!} + \frac{A^3}{3!} + \dots
\end{equation}
is truncated at an order that optimally balances computational cost and accuracy, determined from sharp error bounds based on estimates of $\|A^k\|_1^{1/k}$ for small $k$, with the matrix 1-norm $\|\cdot\|_1$ estimated efficiently.

A key practical advantage of this approach is its \textit{numerical stability}:  
the wave function is propagated through successive sparse matrix-vector multiplications, remaining stable even for extremely small amplitudes.  
By contrast, the split-operator (Trotter) method alternates between real and momentum space, accumulating rounding errors of order $10^{-16}$ per operation in double precision.  
The scaling-and-squaring algorithm performs the evolution in a single backward-stable step, resolving amplitudes as small as the threshold of double precision.
This capability is essential for capturing rare-event fluctuations and extreme-value statistics in the KPZ physics. Consequently, this method enables us to perform simulations of wave packet dynamics under a wide range of disorder strengths in 2D, providing the resolution necessary to probe the universal fluctuation of Anderson localization without resorting to higher-precision arithmetic or excessive computational cost.

\begin{figure}
    \centering
    \includegraphics[width=\linewidth]{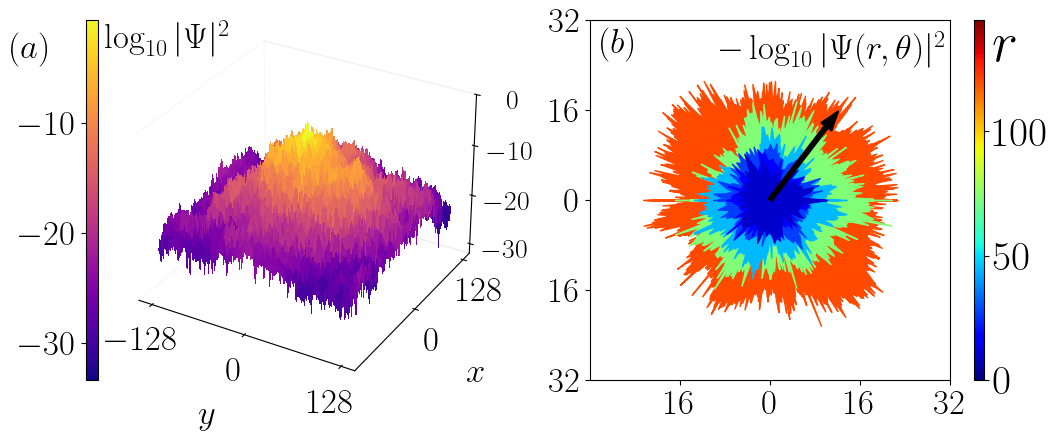}
    \caption{Spatial density of a localized eigenstate and its representation as a rough interface growing with $r$.
(a)~Logarithmic density of an exponentially localized eigenstate, $\ln |\Psi_\alpha(\mathbf{r})|^2$, shown as a 3D surface plot versus $\mathbf{r}=(x,y)$.
(b)~Illustration of the correspondence between 2D Anderson localization and the growth of an effective rough interface.
The radial distance $r = |\mathbf{r}-\mathbf{r}_0|$ from the localization center $\mathbf{r}
_0$, corresponding to the KPZ “time,” is indicated by the color map.
For a fixed radius around the localization center (fixed color), panel~(b) displays the logarithmic density $-\ln |\Psi_\alpha(r,\theta)|^2$ as the distance from the plot center.
Numerical exact diagonalization is performed on a square lattice of size $N = 256 \times 256$ with disorder strength $W=14$.}
    \label{fig:Eigenstates_Representation_InterfaceLike}
\end{figure}

\section{KPZ proporties of 2D localized eigenstates}
\label{sec:Eigenstates}

In this section, we focus on eigenstates of the Hamiltonian in Eq.~\eqref{H_AL} for 2D Anderson localization, computed using the numerical methods described in Sec.~\ref{eigenstate_computation}.
We concentrate on eigenstates near the band center, where both the density of states and the localization length are approximately energy independent.
For each eigenstate, the coordinate origin is chosen at its localization center $\boldsymbol{r}_0$, and we analyze the fluctuations of the logarithmic density $\ln |\Psi(\boldsymbol{r})|^2$ as a function of the radial distance $r$ from $\boldsymbol{r}_0$.

We begin by examining the spatial profile of an exponentially localized eigenstate, shown in Fig.~\ref{fig:Eigenstates_Representation_InterfaceLike}(a). Panel (b) provides an alternative representation emphasizing its analogy with a growing rough interface.
In this correspondence, the logarithmic density plays the role of the interface height, while the radial distance $ r $ from the localization center $\boldsymbol{r}_0$ serves as the effective "time" in the growth process. The polar angle $\theta$ then acts as the spatial coordinate of the interface, as illustrated by the colormap in panel~(b).

Eigenstates display exponential localization away from the center, with the inverse localization length $1/\xi$ corresponding to the velocity of the effective growing interface.
Accordingly, we expect (see also Refs.~\cite{mu_kardar-parisi-zhang_2024,somoza_universal_2007})
\begin{equation}
\ln|\Psi(\boldsymbol{r})|^2 \underset{r \gg \xi}{\approx}
-\frac{2r}{\xi} + \left(\frac{r}{\xi}\right)^{\beta} \Gamma\,\chi(\boldsymbol{r}) + \Lambda,
\label{al_eigen_kpz}
\end{equation}
where $|\Psi(\boldsymbol{r})|^2$ denotes density, $\beta$ is the KPZ fluctuation growth exponent ($\beta=1/3$ in $(1+1)$ dimensions), $\chi(\boldsymbol{r})$ is a random variable of order one, and $\Gamma$ and $\Lambda$ are constants of order unity.
The first term on the right-hand side accounts for the exponential decay of the eigenstate, while the second term describes the fluctuations scaling as $r^{\beta}$.

Eigenstates correspond to the circular version of the interface growth problem, since localization occurs isotropically from the center $\boldsymbol{r}_0$.
This configuration is analogous to the point-initial condition of wave packet dynamics [see Eq.~\eqref{Eq:inicirc}].

For a fixed radius $r$ (i.e., fixed KPZ time) around the localization center, Fig.~\ref{fig:Eigenstates_Representation_InterfaceLike}(b) displays the logarithmic density $\ln \vert \Psi(r,\theta)\vert^2$, represented as the radial distance from the plot center.
This visualization, similar to Fig.~1 of Ref.~\cite{mu_kardar-parisi-zhang_2024} for localized wave packets, strikingly resembles a growing rough interface, see e.g.~\cite{takeuchi_universal_2010}.
In the following, we perform a quantitative analysis of the spatial fluctuations of eigenstates to assess that they belong to the KPZ universality.

\subsection{Fluctuation Scaling}
\begin{figure*}
    \centering
    \includegraphics[width=\textwidth]{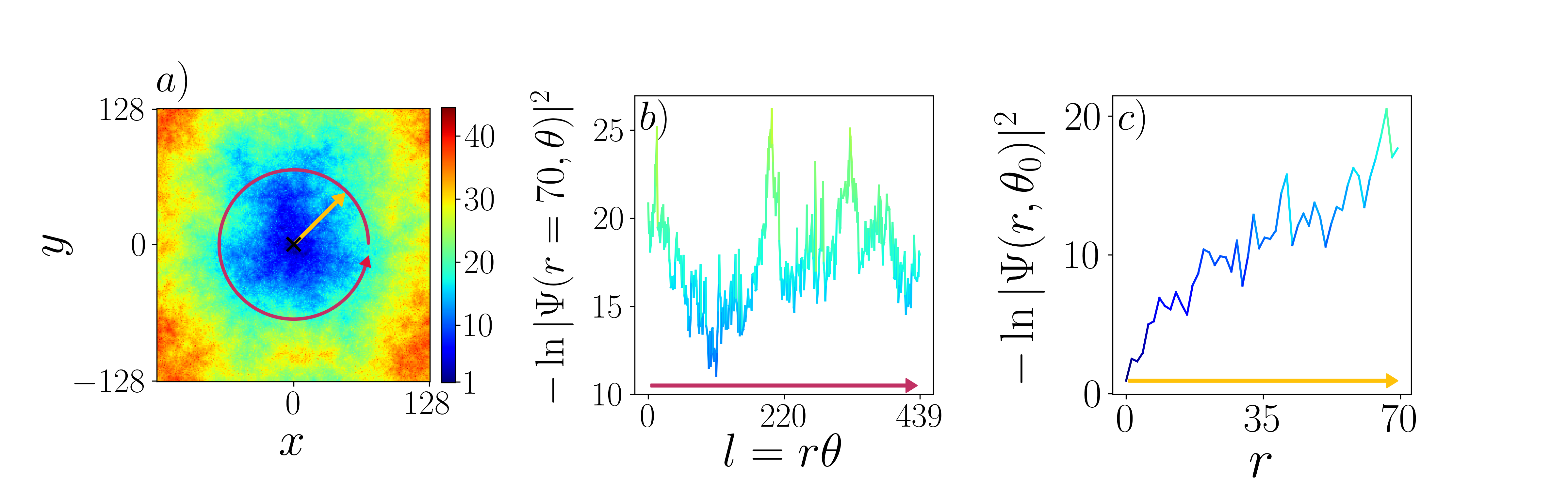}
    \caption{Spatial fluctuations of an eigenstate in 2D Anderson localization.
(a)~Colormap of the logarithmic density of a localized eigenstate, $-\ln |\Psi_\alpha(\boldsymbol{r})|^2$.
The black cross marks the localization center $\boldsymbol{r}_0$, chosen as the lattice center.
The yellow arrow indicates the direction of the diagonal $r$, playing the role of the ``time'' in the KPZ physics for fixed ``spatial'' coordinate $\theta=\pi/4$.
Conversely, the red circle of radius $r=50$ defines the ``spatial'' coordinate $\theta$ at a fixed effective ``time'' $r=50$.
(b)~Fluctuations of the logarithmic density along the circle of radius $r=50$, i.e., $-\ln |\Psi_\alpha(r=50,\theta)|^2$ plotted as a function of $\theta$.
(c)~Logarithmic density along the diagonal direction, $-\ln |\Psi_\alpha(r,\theta=\pi/4)|^2$, as a function of $r$.
The linear growth reflects the exponential localization of the eigenstate, while the fluctuations around this linear trend will be analyzed in Fig.~\ref{fig:Eigenstates_exponentBeta}.
Simulations were performed on a square lattice of size $N = 256\times 256$ with disorder strength $W=10$.
The eigenstate shown lies near the band center at energy $E \approx 0$.
    }
    \label{fig:Eigenstates_VisualisationMethodsBeta}
\end{figure*}

We employ two complementary methods to extract the scaling of fluctuations with distance.  
First, we compute the fluctuations across all angular directions at a fixed radius $r$, corresponding to averaging over circles centered at the localization center. This approach captures fluctuations over the full effective interface in the KPZ analogy.  
Second, we analyze fluctuations along a fixed spatial direction, i.e., at a fixed polar angle $\theta_0$, corresponding to the interface height at a single spatial coordinate.

These two procedures are illustrated in Fig.~\ref{fig:Eigenstates_VisualisationMethodsBeta} and described in detail below.  
In the first method, we calculate the angular average and standard deviation of the logarithmic density $\ln |\Psi(r,\theta)|^2$ along circles of radius $r$, as depicted in panels~(a) and~(b).  
Denoting the angular average (over the effective interface) by $\langle \cdot \rangle_\theta$, the standard deviation at radius $r$ is given by  
\begin{equation}
\sigma_{c}(r) = \sqrt{ \langle (\ln |\Psi(r,\theta)|^2)^2 \rangle_{\theta} - \langle \ln |\Psi(r,\theta)|^2 \rangle_{\theta}^2 }.
\label{EQ:Eigenstates_std_r_circle_Formula}
\end{equation}
This quantity is subsequently averaged over eigenstates near the band center and over disorder realizations.  
In the second method, we fix the polar angle $\theta_0 = \pi/4$, as shown by the yellow arrow in Fig.~\ref{fig:Eigenstates_VisualisationMethodsBeta}(a,c), and compute the mean and variance of $-\ln |\Psi(r,\theta_0)|^2$ as a function of $r$, again averaging over eigenstates and disorder realizations.

\begin{figure}
    \centering
    \includegraphics[width=\linewidth]{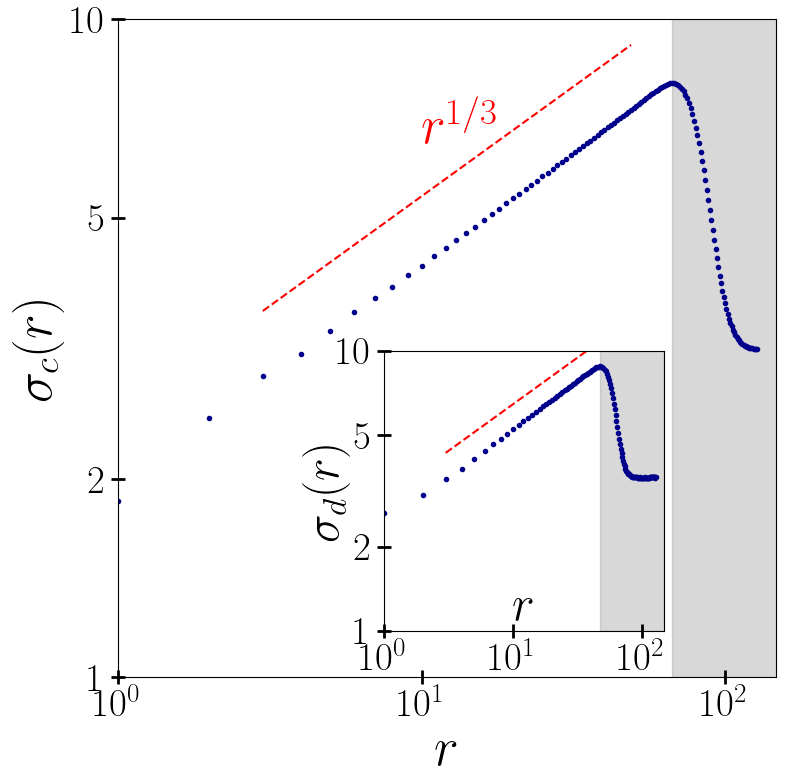}
    \caption{
Standard deviation of the eigenstate logarithmic density (over different eigenstates and disorder configurations) as a function of the distance $r$ from the localization center, displayed on log--log axes.
The main panel shows the fluctuation scaling over circles, $\sigma_c(r)$ [see Eq.~\eqref{EQ:Eigenstates_std_r_circle_Formula}], corresponding to Fig.~\ref{fig:Eigenstates_VisualisationMethodsBeta}(b).
The inset shows the fluctuation scaling along the diagonal direction, $\sigma_d(r)$, corresponding to Fig.~\ref{fig:Eigenstates_VisualisationMethodsBeta}(c).
Both methods exhibit power-law growth with an exponent consistent with the universal KPZ value $1/3$ (red dashed line: $\sigma \sim r^{1/3}$).
The shaded region indicates distances where the eigenstate density falls below numerical precision.
Simulations are performed on a square lattice of size $N = 256 \times 256$ with disorder strength $W = 14$, averaging 300 eigenstates near $E\approx0$ over 400 disorder realizations.
    }
    \label{fig:Eigenstates_exponentBeta}
\end{figure}

Both approaches yield consistent results, as shown in Fig.~\ref{fig:Eigenstates_exponentBeta}.  
They confirm the analogy with KPZ physics, displaying a fluctuation growth exponent $\beta \approx 1/3$, distinctly different from the value $\beta = 1/2$ for 1D Anderson localization.

\begin{figure}[h!]
    \centering
    \includegraphics[width=0.95\linewidth]{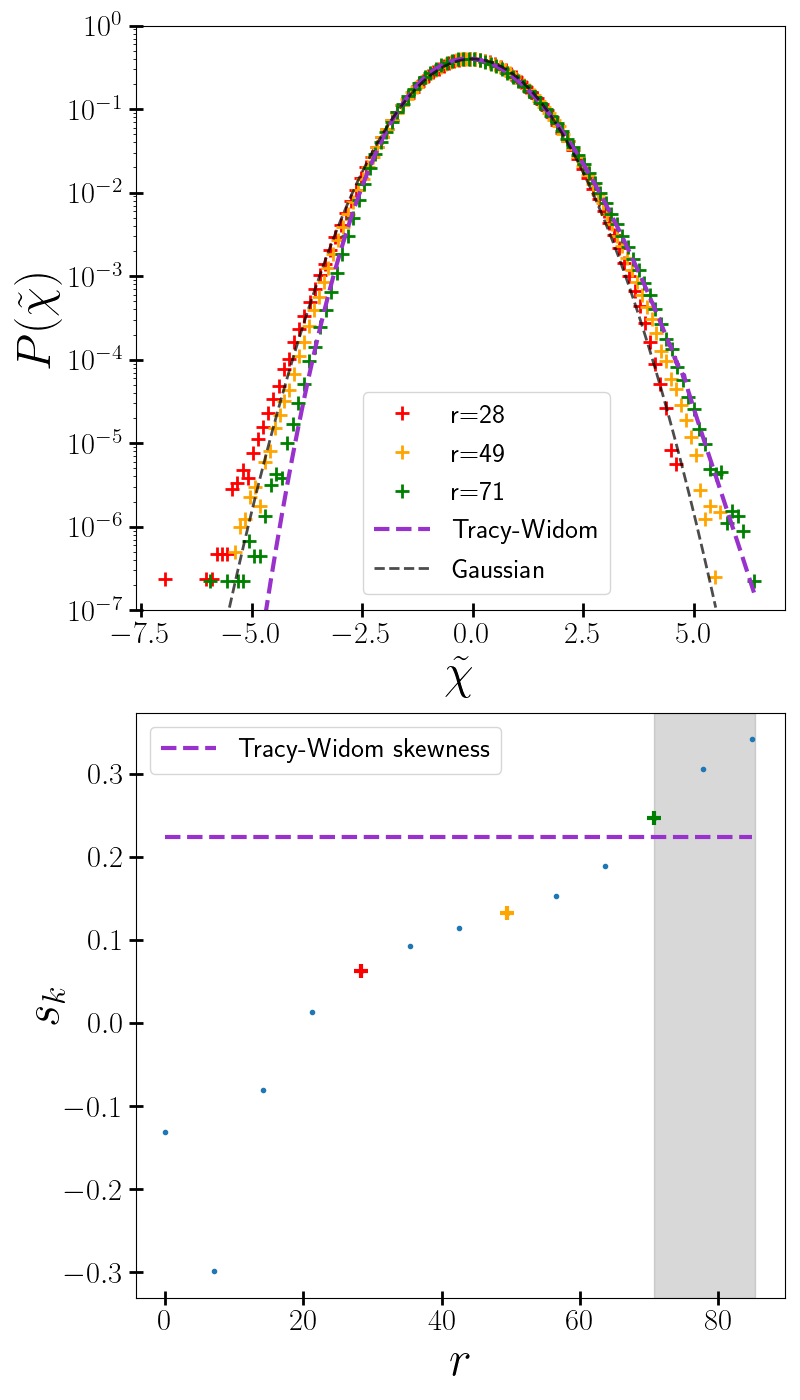}
    \caption{Distribution of the eigenstate logarithmic density and its skewness.
(a)~Probability distribution of $\chi = -\ln|\Psi(r,\theta=\pi/4)|^2$, rescaled by its mean and standard deviation, $\tilde{\chi} = (\chi - \overline{\chi}) / \sigma_\chi$, for three distances $r = 30,\,50,\,70$ (red, orange, and green, respectively).  
As $r$ increases, the distribution crosses over from nearly Gaussian (black dashed line) toward the Tracy-Widom for GUE (purple dashed line).  The reference Tracy-Widom for GUE is obtained using the Python module \texttt{https://github.com/yymao/TracyWidom} and rescaled to zero mean and unit standard deviation.  
(b)~Skewness $s_k(r)$ [Eq.~\eqref{Eq:SkewnessDef}] as a function of $r$, compared with that of the Tracy--Widom for GUE $s_{k_\mathrm{TW_{GUE}}}\approx 0.22$ (purple dotted line).  
The gray region marks distances where the wavefunction density approaches numerical precision limits.  
Simulations are performed on a square lattice of size $N = 128 \times 128$ with disorder strength $W = 13$.}    
    \label{fig:Eigenstates_DistributionTW}
\end{figure}

We verified that the fluctuation growth exponent remains unchanged when the onsite disorder is drawn from a Gaussian distribution instead of a uniform one.  
Furthermore, eigenstates away from the band center exhibit the same fluctuation scaling exponent, consistent with the KPZ universality.

\subsection{Tracy-Widom distribution of the density logarithm of 2D localized eigenstates}

We now examine the probability distribution of the logarithmic density, $\chi = -\ln |\Psi(r,\theta)|^{2}$, measured at various radial distances $r$ from the localization center along a fixed diagonal direction ($\theta = \pi/4$).
Figure~\ref{fig:Eigenstates_DistributionTW}(a) shows that, after rescaling by the mean $\overline{\chi}$ and variance $\sigma_\chi^2 = \overline{\chi^2} - \overline{\chi}^2$, the distributions of the normalized variable $\tilde{\chi} = (\chi - \overline{\chi}) / \sigma_\chi$ collapse onto the Tracy-Widom distribution for the Gaussian Unitary Ensemble (GUE).
This provides strong evidence that the fluctuations of 2D localized eigenstates belong to the KPZ universality class.

Unlike a Gaussian distribution, the Tracy-Widom distribution is asymmetric and thus characterized by a nonzero skewness,
\begin{equation}
s_k = \frac{\sum_i (x_i - \bar{x})^3 p_i}{\sigma^{3}},
\label{Eq:SkewnessDef}
\end{equation}
where $x_i$ are the binned values of $\ln |\Psi|^{2}$, $p_i$ the corresponding probabilities, and $\sigma$ the standard deviation.
The evolution of the skewness with distance $r$ is shown in Fig.~\ref{fig:Eigenstates_DistributionTW}(b), where it is seen to converge toward the value of the Tracy-Widom GUE distribution.
Our numerical results approach this asymptotic value for sufficiently large $r$, with deviations at larger distances arising from the limits of numerical precision.

\section{Microscopic structure of 2D localized states: dominant paths and their glassy properties}
\label{sec:dominant_paths}

Beyond the KPZ universality class that governs the spatial fluctuations of eigenstates in their localized tails, an additional layer of \textit{glassy physics} emerges in the form of dominant paths, which determine the microscopic organization of individual eigenstates.  
The central question addressed in this section is how a single eigenstate localizes spatially within a given disorder realization, and how this structure relates to the glassy physics of directed polymers, an archetypal system of pinned elastic manifolds exhibiting glassy features such as pinning and avalanches~\cite{mezard_glassy_1990,halpin-healy_kinetic_1995}.  
Such dominant paths and their glassy properties have been identified in previous studies of quantum conductance fluctuations~\cite{markos_electron_2010,lemarie_glassy_2019} and in the long-time evolution of localized wave packets in 2D Anderson insulators~\cite{mu_kardar-parisi-zhang_2024}.  

To probe the spatial structure of a localized eigenstate, we employ a method inspired by Refs.~\cite{lemarie_glassy_2019,mu_kardar-parisi-zhang_2024}, conceptually analogous to the scanning-gate microscopy.
A small local perturbation is introduced at a site $\boldsymbol{r}'$ in the form of a weak modification of the on-site disorder potential, and we study how this perturbation alters the density of an eigenstate at a distant observation point $\boldsymbol{r}_{\mathrm{obs}}$, located well beyond the localization length, $|\boldsymbol{r}_{\mathrm{obs}} - \boldsymbol{r}_0| \gg \xi$.
The unperturbed eigenstate is denoted by $\ket{\Psi_n^{(0)}}$, localized around $\boldsymbol{r}_0$, with amplitude $\Psi_n^{(0)}(\boldsymbol{r}_0)$ of order unity; the superscript $(0)$ indicates unperturbed quantities.

Within first-order perturbation theory, the perturbed eigenstate can be written as
\begin{equation}
\label{Eq:Eigenstates_perturbation_theory_State}
\ket{\Psi_n^{(1)}} = \ket{\Psi_n^{(0)}} 
+ \sum_{\alpha \neq n} 
\frac{\bra{\Psi_\alpha^{(0)}} V_{\boldsymbol{r}'} \ket{\Psi_n^{(0)}}}
{E_n - E_\alpha}
\ket{\Psi_\alpha^{(0)}},
\end{equation}
where the sum runs over all unperturbed eigenstates $\ket{\Psi_\alpha^{(0)}}$ of the Hamiltonian~\eqref{H_AL}.  
The local perturbation is modeled as $V_{\boldsymbol{r}'} = \epsilon_{\boldsymbol{r}'} \delta(\boldsymbol{r}-\boldsymbol{r}')$, with amplitude $\epsilon_{\boldsymbol{r}'} \ll w_{\boldsymbol{r}'}$ (typically less than $0.01\%$ of the local disorder strength).  
The corresponding perturbed wavefunction at site $\boldsymbol{r}_{\mathrm{obs}}$ is then given by
\begin{eqnarray}
\label{Eq:Eigenstates_perturbation_theory_Wave}
\Psi_n^{(1),\boldsymbol{r}'}(\boldsymbol{r}_{\mathrm{obs}}) 
&= &\Psi_n^{(0)}(\boldsymbol{r}_{\mathrm{obs}}) \nonumber \\
&+& \sum_{\alpha \neq n} 
\epsilon_{\boldsymbol{r}'} 
\frac{
\Psi_\alpha^{(0)\ast}(\boldsymbol{r}')
\Psi_n^{(0)}(\boldsymbol{r}')
}{
E_n - E_\alpha
}
\Psi_\alpha^{(0)}(\boldsymbol{r}_{\mathrm{obs}}).\nonumber \\
\end{eqnarray}

To quantify how a local perturbation at $\boldsymbol{r}'$ affects the amplitude of an eigenstate at position $\boldsymbol{r}_{\mathrm{obs}}$, we define a response function as the relative change between the perturbed and unperturbed eigenstate densities at $\boldsymbol{r}_{\mathrm{obs}}$, normalized by the local perturbation strength:
\begin{equation}
\label{Eq:Eigenstates_ResponseFunction}
\rho_{\boldsymbol{r}_{\mathrm{obs}}}(\boldsymbol{r}') 
= \frac{1}{\epsilon(\boldsymbol{r}')}
\frac{\left|\, |\Psi_n^{(1),\boldsymbol{r}'}(\boldsymbol{r}_{\mathrm{obs}})|^2 
- |\Psi_n^{(0)}(\boldsymbol{r}_{\mathrm{obs}})|^2 \,\right|}
{|\Psi_n^{(0)}(\boldsymbol{r}_{\mathrm{obs}})|^2}.
\end{equation}
This response function quantifies how sensitive the local eigenstate amplitude at $\boldsymbol{r}_{\mathrm{obs}}$ is to weak perturbations applied elsewhere in the system.

In principle, evaluating $\rho_{\boldsymbol{r}_{\mathrm{obs}}}(\boldsymbol{r}')$ from first-order perturbation theory requires knowledge of the full eigensystem of the Hamiltonian, which demands complete diagonalization and thus limits the accessible system size.  
To overcome this limitation, we benchmarked the minimal number of eigenstates necessary to faithfully reproduce the full-spectrum response.  
We find that including a sufficient number of eigenstates yields a response that coincides with that obtained from full diagonalization.  
This allows us to employ sparse diagonalization techniques.  
In practice, we find that using $\approx 10^3$ eigenstates for system sizes of order $N = 256 \times 256 = 65536$ gives reliable results.

\begin{figure*}
    \centering
8    \includegraphics[width=\textwidth]{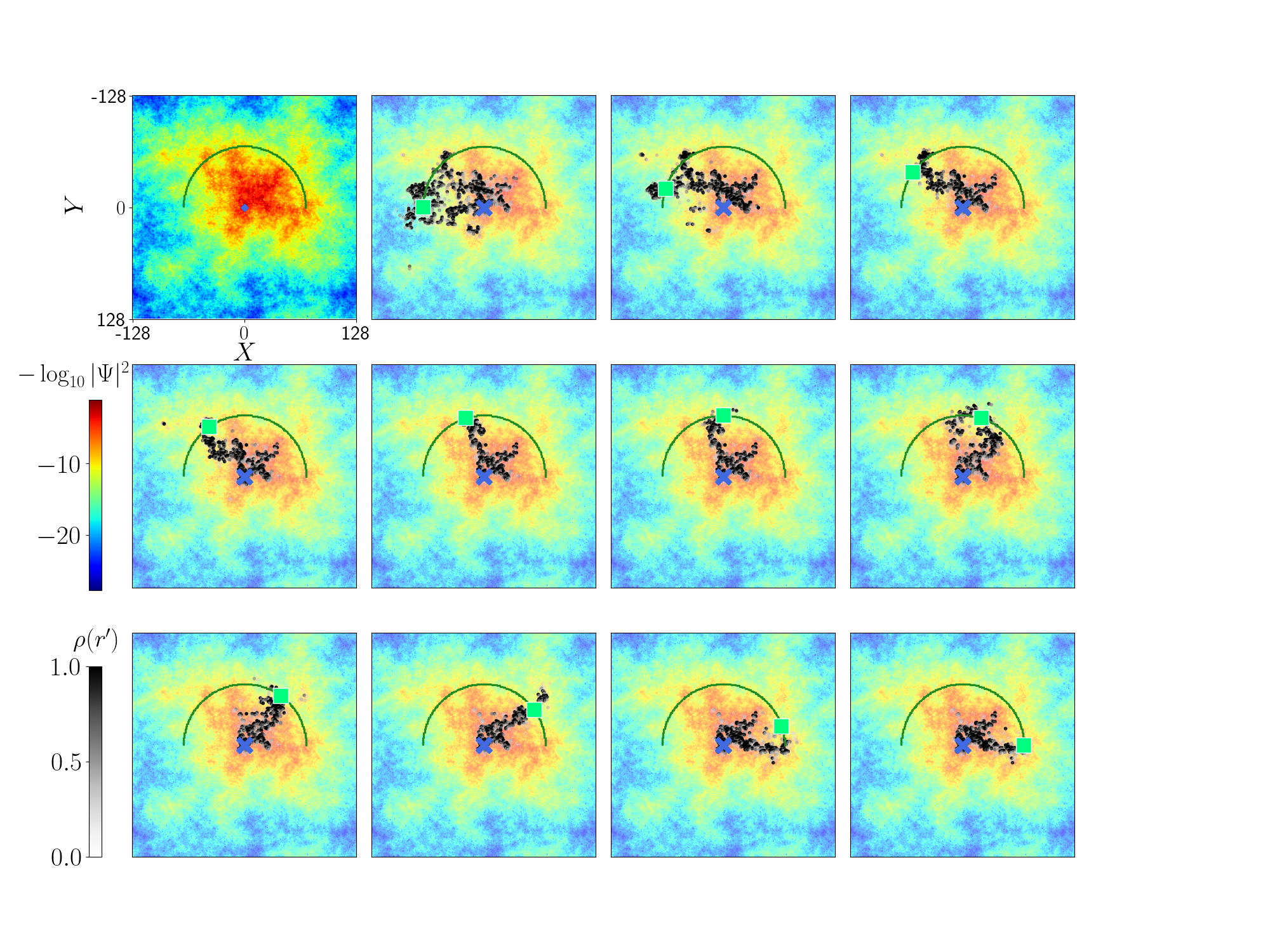}
    \caption{Microscopic structure of a localized eigenstate and dominant paths for different observation points.
Top left panel: colormap of the logarithmic density of a localized eigenstate, $\ln|\Psi(\boldsymbol{r})|^2$. 
The blue cross marks the localization center, and the green semicircle has radius $r = 70$. 
We consider eleven evenly spaced observation sites $\boldsymbol{r}_{\mathrm{obs}}$ on this semicircle, for which we compute the response function $\rho_{\boldsymbol{r}_{\mathrm{obs}}}(r')$ [see Eq.~\eqref{Eq:Eigenstates_ResponseFunction}]. 
Each response function is displayed in a separate panel in grayscale with transparency, overlaid on the eigenstate density.
Resonances can occur at specific sites along the paths, producing arbitrarily large response values and obscuring the underlying structure, hence we restrict the plotted response values to the range $[0,1]$. Most panels exhibit a sharply concentrated response along a single path connecting the localization center to the observation point, strongly reminiscent of pinned directed-polymer configurations. 
Other panels display a more diffuse response; we show in the next section that these correspond to abrupt switches between pinned configurations, analogous to avalanches in glassy systems. 
We also observe that pinned paths follow the anisotropy of the eigenstate density, indicating that the anisotropic localization structure can be interpreted in terms of dominant paths.
Simulations are performed on a square lattice of size $N=256\times 256$ with disorder strength $W=9$. 
The shown eigenstate lies near the band center, and 2000 neighboring eigenstates in energy are used to construct the perturbed state entering Eq.~\eqref{Eq:Eigenstates_ResponseFunction}.
    }
    \label{fig:PathsPinning_GifFrames}
\end{figure*}

\subsection{Dominant paths and the microscopic structure of a localized eigenstate}
\label{subsec:dominant_paths_structure}

We compute the response function $\rho_{\boldsymbol{r}_{\mathrm{obs}}}(\boldsymbol{r}')$ for a fixed observation point $\boldsymbol{r}_{\mathrm{obs}}$, while applying local perturbations of the disorder potential at different sites $\boldsymbol{r}'$.
The resulting response maps for a representative localized eigenstate are shown in Fig.~\ref{fig:PathsPinning_GifFrames}.
Specifically, the upper-left panel displays the logarithm of the eigenstate density as a color map, while all other panels present the corresponding response function $\rho_{\boldsymbol{r}_{\mathrm{obs}}}(\boldsymbol{r}')$ for different choices of $\boldsymbol{r}_{\mathrm{obs}}$.
Each response map (shown in grayscale) is superimposed on the logarithmic density of the same eigenstate to highlight spatial correlations between the two.

For a given observation point $\boldsymbol{r}_{\mathrm{obs}}$, the response is strongly inhomogeneous and exhibits enhanced sensitivity along a distinct path connecting the localization center (starting point) to the observation site $\boldsymbol{r}_{\mathrm{obs}}$ (end point).
Because a few local resonances may occur at isolated sites along these paths, we restrict the displayed response values to the range $[0,1]$ for clarity.
We have checked that restricted and unrestricted response maps give the same dominant paths.

When the observation point $\boldsymbol{r}_{\mathrm{obs}}$ is varied along the green circle in Fig.~\ref{fig:PathsPinning_GifFrames}, that is at a fixed distance $|\boldsymbol{r}_{\mathrm{obs}} - \boldsymbol{r}_0| \gg \xi$ from the localization center, we obtain a set of distinct dominant paths.  
The correspondence between these dominant paths and the microscopic structure of the eigenstate, characterized by its logarithmic density, constitutes a first key result.  
We observe that the eigenstate density does not decay isotropically from the localization center: certain directions form branches of weaker localization, where the density decays more slowly.  
When $\boldsymbol{r}_{\mathrm{obs}}$ lies along such a branch, the dominant path strongly overlaps with the branch, as illustrated in the two upper-right panels of Fig.~\ref{fig:PathsPinning_GifFrames}.  
Conversely, when the observation point lies between two branches, i.e., in regions of faster decay and lower eigenstate density, the dominant path broadens substantially, as seen in the upper-left panels.  

These observations demonstrate that the dominant paths revealed by the response function faithfully trace the anisotropic structure of individual eigenstates.  
In the following, we interpret these findings within the framework of glassy physics of directed polymers, which provides a natural description underlying such spatial organization.

\subsection{Pinning and Avalanches}

\begin{figure}
    \centering
    \includegraphics[width=.48\textwidth]{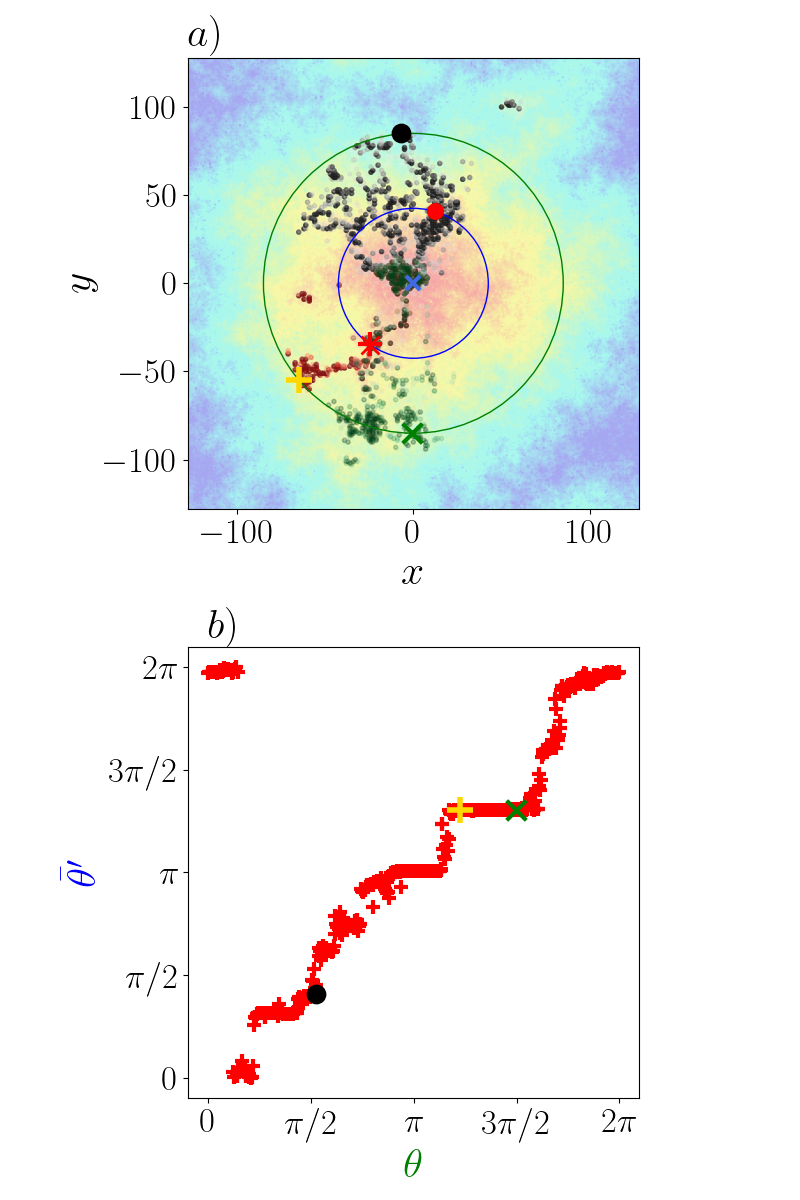}
    \caption{ Pinning and avalanche behavior of the dominant paths of eigenstates in 2D Anderson localization.
(a)~Illustration of the procedure used to characterize pinning and avalanches of dominant paths for a localized eigenstate, whose probability density is shown in logarithmic scale. 
The outer green circle represents the set of observation points $\boldsymbol{r}_{\mathrm{obs}}$ at radius $r = |\boldsymbol{r}_{\mathrm{obs}} - \boldsymbol{r}_0| = 85$ and polar angle~$\theta$. 
For each observation point, we compute the response function and extract its average angular position $\overline{\theta'}$ along the blue circle of radius $r/2$ [see Eq.~\eqref{Eq:Eigenstates_Paths_AveragePosition}]. 
(b)~Average angular position $\overline{\theta'}$ as a function of the observation angle~$\theta$. 
Plateaus correspond to situations where the dominant path at mid-distance remains unchanged as $\theta$ varies, indicating pinning. 
Panel~(a) shows examples of dominant paths for two observation points lying on the same plateau around $\theta \sim 5\pi/4$ (marked by the yellow horizontal cross and the green diagonal cross). 
Abrupt jumps between plateaus signal avalanche events, where the response function can be interpreted as a superposition of several competing dominant paths. 
Panel~(a) highlights the broad response function obtained for the black circular observation point located in an avalanche region around $\theta \sim \pi/2$.
    Numerical simulations correspond to system size $N=256 \cross 256$ and disorder strength $W=11$, with the considered eigenstate in the band center near energy $E\approx0$.}
    \label{fig:Eigenvectors_Paths_PinningAvalanches}
\end{figure}

A key hallmark of glassy behavior in disordered systems is the occurrence of pinning and avalanches~\cite{mezard_glassy_1990,halpin-healy_kinetic_1995,jogi_self-organized_1998,thiery_analytical_2017}.
The directed polymer (DP) problem provides a paradigmatic example of such behaviors~\cite{calabrese_free-energy_2010,sales_fragility_2002}.
In $(1+1)$ dimensions, a polymer extends in one direction, conventionally referred as "time" $t$, while it can fluctuate along the transverse spatial direction $x$. The polymer experiences a quenched random potential $V(x,t)$, and the energy of a given configuration corresponds to the sum of the on-site potential energies along the path.
At zero temperature, the system minimizes this total energy; at finite temperature, one introduces Boltzmann weights, leading to a partition function and free energy landscape.

In $(1+1)$ dimensions, the directed polymer is always in a glassy phase governed by a zero-temperature fixed point, where thermal fluctuations are suppressed and the configuration is dominated by the disordered energy landscape~\cite{mezard_glassy_1990,sales_fragility_2002}.
In this regime, the polymer becomes pinned in a globally optimal configuration determined by the specific realization of disorder, the so-called dominant path.
When a control parameter or the polymer endpoint is varied, the dominant configuration often remains unchanged (pinning).
However, beyond certain thresholds, the polymer may abruptly rearrange into a distinct configuration far from the previous one, a.k.a. an avalanche.

In the following, we demonstrate that analogous behavior emerges in the dominant paths extracted from eigenstates of 2D Anderson localization.
As shown in Fig.~\ref{fig:Eigenvectors_Paths_PinningAvalanches}(a), when the observation point $\boldsymbol{r}_{\mathrm{obs}}$ is displaced, the dominant path typically remains unchanged, indicating a pinned configuration. Upon further displacement, however, we observe sudden structural rearrangements of the path, and these are direct analogs of avalanches.

We analyze a localized eigenstate centered at $\boldsymbol{r}_0$ taken as the origin of coordinates to quantify this behavior.
We consider a set of observation points $\boldsymbol{r}_{\mathrm{obs}}$ lying on the green circle in Fig.~\ref{fig:Eigenvectors_Paths_PinningAvalanches}(a), at a fixed distance $r = |\boldsymbol{r}_{\mathrm{obs}} - \boldsymbol{r}_0|$, parameterized as $(r,\theta)$.
For each angle $\theta$, we compute the response function $\rho_{\boldsymbol{r}_{\mathrm{obs}}}(\boldsymbol{r}')$ and analyze the angular distribution of the response at an intermediate radius $r' = r/2$, effectively probing the path geometry halfway between the localization center and the observation point.
We define the average angular position of the path on this midpoint circle as
\begin{equation}
\overline{\theta'} =
\arctan \left(
\frac{\sum_i \sin(\theta'_i)\,\tilde{\rho}(\theta'_i)}
{\sum_i \cos(\theta'_i)\,\tilde{\rho}(\theta'_i)}
\right),
\label{Eq:Eigenstates_Paths_AveragePosition}
\end{equation}
where $\tilde{\rho}(\theta'_i) = \rho_{\boldsymbol{r}_{\mathrm{obs}}}(r/2,\theta'_i) / \sum_j \rho_{\boldsymbol{r}_{\mathrm{obs}}}(r/2,\theta'_j)$ is the normalized response on the blue circle (see Fig.~\ref{fig:Eigenvectors_Paths_PinningAvalanches}(a)).
Numerically, the discrete sums are over lattice sites forming an approximate circle of radius $r/2$.

This construction is illustrated in Fig.~\ref{fig:Eigenvectors_Paths_PinningAvalanches}(a) for three different endpoints on the green circle.
The average angular positions on the midpoint circle, marked by red symbols, trace the mean trajectory of each path.
Sweeping the observation angle $\theta$ effectively probes the internal structure of the localized eigenstate in all directions.
We observe distinct, well-defined branches (e.g., the green/$\times$ and yellow/$+$ paths) separated by abrupt jumps (black/$\circ$ path), directly reminiscent of pinning and avalanches in directed polymers~\cite{lemarie_glassy_2019}.
In Fig.~\ref{fig:Eigenvectors_Paths_PinningAvalanches}(b), we quantify this behavior through the midpoint angle $\overline{\theta'}$ plotted as a function of the observation angle $\theta$.
Plateaus correspond to pinned configurations where the dominant path remains stable as $\theta$ varies, while sharp jumps mark rearrangements akin to avalanches where competing paths exchange dominance.
The emergence of such discontinuous transitions provides clear evidence of glassy behavior in the microscopic structure of eigenstates in 2D Anderson localization.

\subsection{Dominant path scaling}
As shown in the previous section, the transverse position of the dominant path exhibits plateau behavior as a function of the endpoint angle, with abrupt jumps separating successive plateaus. We interpret these features as pinning and avalanche, characteristic hallmarks of glassy physics in directed polymers and, more broadly, in pinned elastic manifolds.
Here, we investigate how this behavior depends on the path length and whether it follows the same scaling as directed polymer.  
In particular, the typical jump size between plateaus is expected to grow as a power law of the path length, with exponent $\zeta = 2/3$, the wandering exponent~\cite{halpin-healy_kinetic_1995}, which is directly related to the correlation length behavior along a KPZ interface.

\begin{figure}
    \centering
    \includegraphics[width=\linewidth]{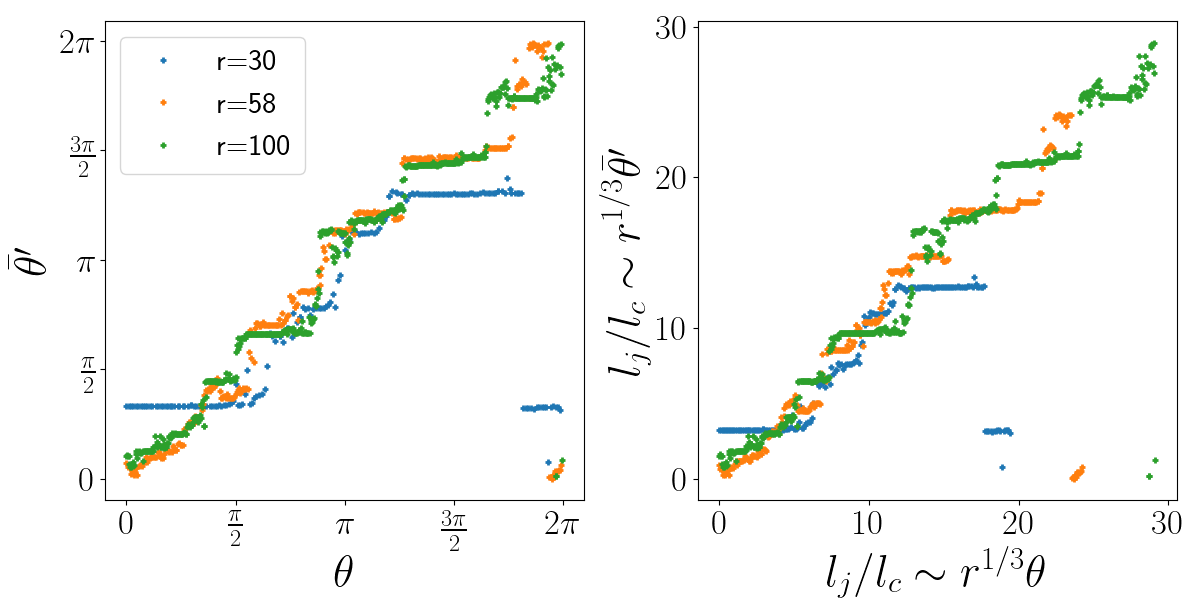}
    \caption{Pinning and avalanches in 2D Anderson localization: dependence on path length. 
(a)~Average angular position $\overline{\theta'}$ [Eq.~\eqref{Eq:Eigenstates_Paths_AveragePosition}] along the mid-distance circle as a function of the observation point angle $\theta$, shown for three radii $r=30,58,100$. 
As $r$ increases, plateaus fragment into smaller ones separated by smaller jumps, indicating that dominant paths branch as their length grows. 
(b)~Rescaled data comparing the arc-length is $l=r\theta$ with the correlation length $l_c \sim r^\zeta$ with $\zeta = 2/3$, inferred from the scaling of directed polymer. 
Under this rescaling, the plateau widths and jump sizes become approximately independent of $r$. 
This qualitative picture is complemented by a quantitative analysis of the standard deviation of the mid-point position over disorder realizations, shown in Fig.~\ref{fig:eigen_roughnessPaths}. 
Simulations are performed on a square lattice of size $N=256\times256$ with disorder strength $W=11$, using 2000 eigenstates near the band center ($E \approx 0$) in the perturbative calculation.
}
    \label{fig:rdependencyofpinning}
\end{figure}

We illustrate how increasing the endpoint distance $r$ affects the pinning and avalanche behavior of the dominant paths in Fig.~\ref{fig:rdependencyofpinning}.  
We show that, upon rescaling the arc length using the appropriate correlation scale $l_c \sim r^{\zeta} = r^{2/3}$, both the plateau widths and the jump magnitudes collapse to approximately the same scale for all values of $r$.

More quantitatively, we compute the response function $\rho$ for observation points located at distance $r$ along the diagonal direction, i.e., $\theta = \pi/4$.  
We then extract the transverse position of the path at mid-distance $r/2$ along the anti-diagonal by evaluating
\begin{equation}
\label{Eq:Eigenstates_Paths_AveragePosition}
\bar{v} = \sum_{i} v_i \, \overline{\rho}(v_i),
\end{equation}
where $v_i$ labels the lattice sites along the anti-diagonal at mid-distance, and
\begin{equation}
\overline{\rho}(v_i)
= 
\frac{\rho_{\boldsymbol{r}_{\mathrm{obs}}}(r/2, v_i)}
{\sum_j \rho_{\boldsymbol{r}_{\mathrm{obs}}}(r/2, v_j)}
\end{equation}
is the normalized response on this line, which can be interpreted as a probability distribution.

\begin{figure}
    \centering
    \includegraphics[width=.48\textwidth]{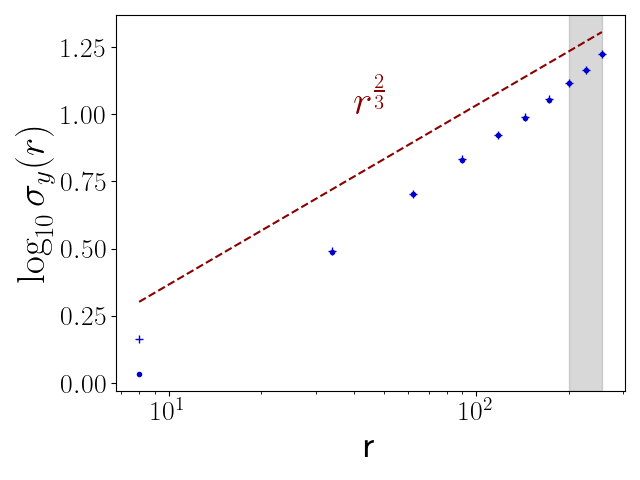}
    \caption{Wandering of the dominant paths in the diagonal geometry. 
Variance of the path position at mid-distance, computed from Eq.~\eqref{Eq:Eigenstates_Paths_AveragePosition}, averaged over eigenstates and disorder realizations. 
The variance is shown on a logarithmic scale and exhibits an approximately linear behavior, indicating power-law scaling. 
The observed exponent is compatible with the wandering exponent $2/3$ of directed polymer, illustrated by the red dashed line. 
Simulations are performed on a square lattice of size $N = 256 \times 256$ with disorder strength $W = 11$. 
Perturbation theory includes between 1000 and 2000 eigenstates near the band center ($E \approx 0$), with no significant change in the results.
    }
    \label{fig:eigen_roughnessPaths}
\end{figure}

The quantity $\bar{v}$ therefore represents the average transverse displacement of the path at mid-distance.  
We compute its distribution over disorder realizations and extract its standard deviation $\sigma_{\bar{v}}$ as a function of $r$.  
As shown in Fig.~\ref{fig:eigen_roughnessPaths}, the results are fully consistent with the scaling of directed polymer:
\begin{equation}
\sigma_{\bar{v}} \sim r^{\zeta},
\end{equation}
with an exponent close to $2/3$, the wandering exponent of the $(1\!+\!1)$-dimensional directed polymer, equivalently the inverse dynamic exponent of the KPZ universality class.

\section{Stretched-exponential form of 2D localized wave packets and single-parameter scaling}
\label{Analytical_form}

We now address the second objective of this work: characterizing the KPZ scaling of localized wave packets. This extends our previous analysis~\cite{mu_kardar-parisi-zhang_2024} by employing a highly efficient numerical scheme that accurately resolves the exponentially small tails of localized wave packets, thereby reconciling emergent KPZ fluctuations with the single-parameter scaling hypothesis of Anderson localization. Before this, we relate long-time evolved wave packets to eigenstate correlations and demonstrate the KPZ scaling for the latter.

\subsection{Long-time evolved wave packets and eigenstate correlations}

For wave packet dynamics after a quantum quench of the initial condition at arbitrary time $t$, the density at position $\boldsymbol r$ reads
\begin{eqnarray}
|\psi^t(\boldsymbol r)|^2 &=& \left|\sum_n e^{-iE_n t}\, c_n \Psi_n(\boldsymbol r)\right|^2 \nonumber\\
&=& \sum_n |c_n|^2 |\Psi_n(\boldsymbol r)|^2 \nonumber\\
&&+ \sum_{m\neq n} c_m^{*}c_n\, \Psi_m^{*}(\boldsymbol r)\Psi_n(\boldsymbol r)\, e^{-i(E_n - E_m)t},\nonumber\\
\label{eq:eigen_cross}
\end{eqnarray}
where $c_n = \langle \Psi_n | \psi_0 \rangle$ denotes the overlap of the initial wave packet with eigenstate $\Psi_n$. In the long-time limit, the off-diagonal terms in Eq.~\eqref{eq:eigen_cross} generate persistent temporal fluctuations, but the disorder-averaged spatial profile is dominated by the diagonal term.

\begin{figure}
\includegraphics[width=0.85\linewidth]{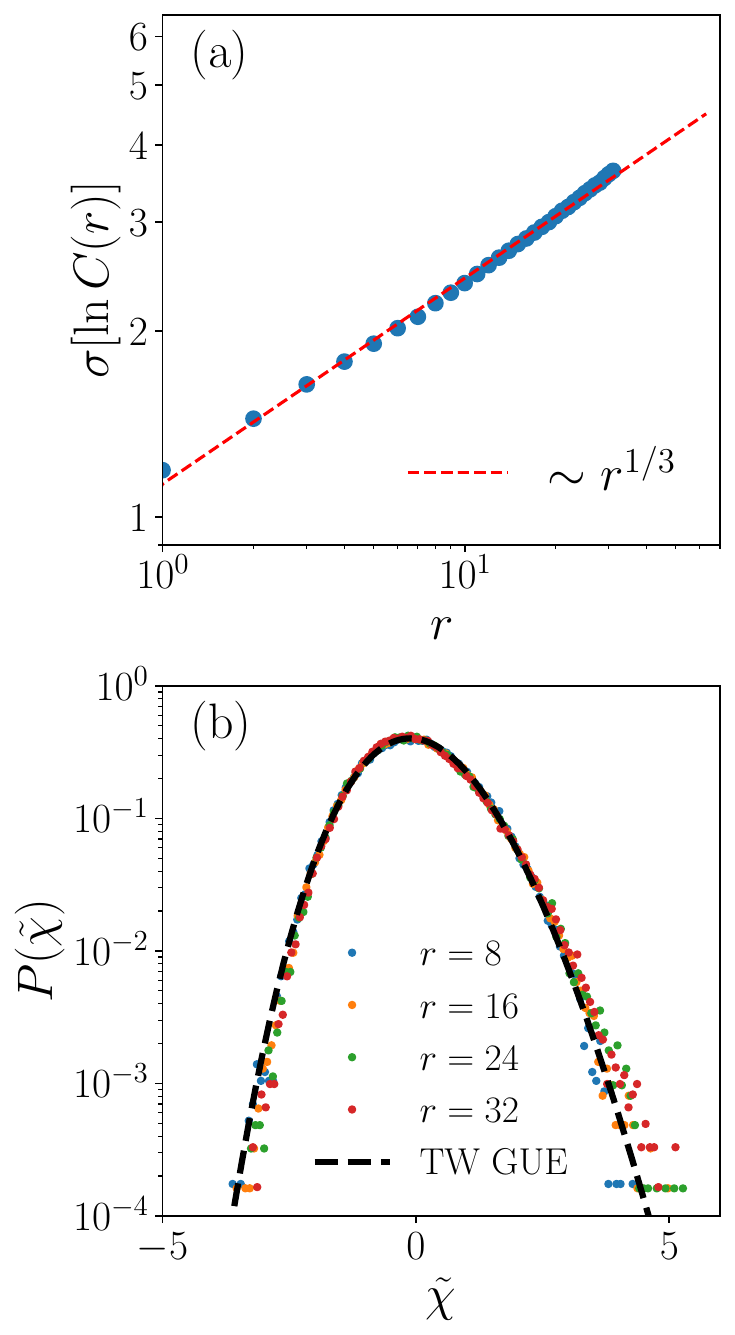}
\caption{Eigenstate correlations in 2D Anderson localization defined in Eq.~\eqref{eigen_corr}. 
(a) Standard deviation of $\ln C(\boldsymbol r)$ as a function of distance $r$. The red dashed line $\propto r^{1/3}$ is a guide to the eye. 
(b) Normalized distribution of the rescaled variable $\tilde\chi(\boldsymbol r)=\big(\ln C(\boldsymbol r)-\langle\ln C(\boldsymbol r)\rangle\big)/\sigma[\ln C(\boldsymbol r)]$ for several distances $r$ (different colors), compared with the rescaled Tracy-Widom distribution for GUE (dashed black line). 
Exact diagonalizations are performed on a square lattice of size $N=64\times64$ under open boundary condition with disorder strength $W=12$ and $7.2\times10^{4}$ disorder realizations.}
\label{fig:anderson_eigen_corr_std_tw}
\end{figure}

For the point initial condition (see Eq.~\eqref{Eq:inicirc}), the long-time wave packet therefore reduces, within the diagonal approximation, to the eigenstate correlation function
\begin{equation}
|\psi^t(\boldsymbol r)|^2 
\underset{t\to\infty} \approx 
C(\boldsymbol r) 
\coloneq \sum_n |\Psi_n(\boldsymbol 0)|^2\,|\Psi_n(\boldsymbol r)|^2.
\label{eigen_corr}
\end{equation}
Since fluctuations of individual eigenstates exhibit the KPZ scaling, as established in the previous sections, we expect $\ln C(\boldsymbol r)$ to display the same phenomenology. This expectation is fully borne out by our numerical results in Fig.~\ref{fig:anderson_eigen_corr_std_tw}, namely the growth of fluctuations is consistent with the KPZ exponent $1/3$ and the fluctuation statistics match the Tracy-Widom distribution for GUE.

\subsection{Proposed expressions for 2D localized wave packets}

We now describe the spatial fluctuation of long-time evolved wave packets obtained using the numerical method discussed in Sec.~\ref{sec:timeevol}. The accuracy and efficiency of this method far exceed the numerical approaches used previously for eigenstates and in Ref.~\cite{mu_kardar-parisi-zhang_2024} for the 2D quantum kicked rotor. In particular, it gives reliable access to the exponentially small tails of the wave packets, allowing us to investigate whether the emergent KPZ scaling is compatible with the single-parameter scaling (SPS), a cornerstone of Anderson localization.

\begin{figure}
\includegraphics[width=1.0\linewidth]{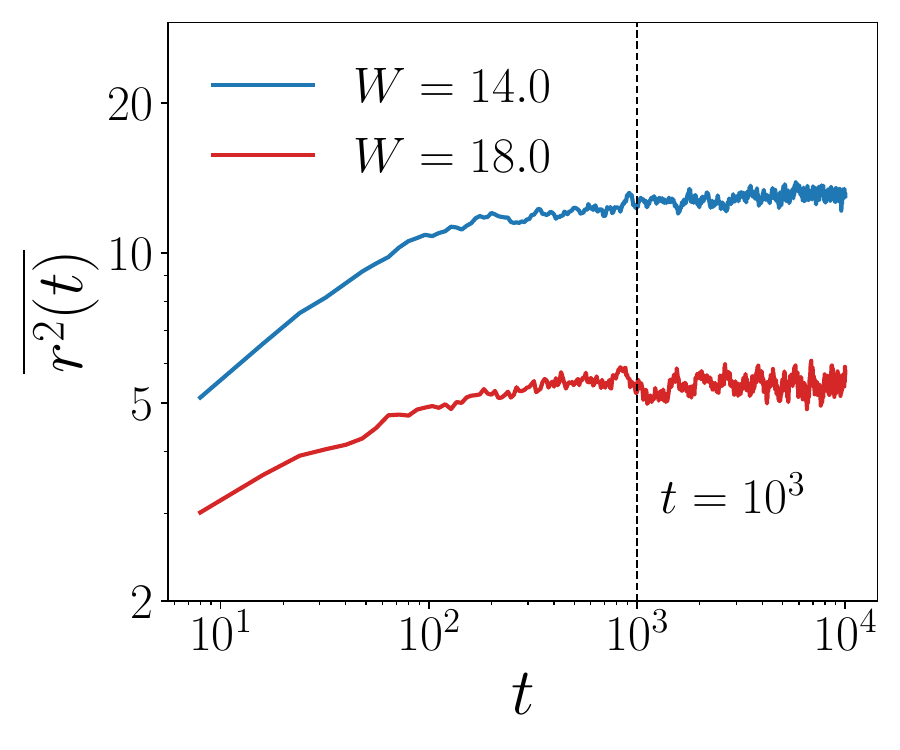}
\caption{Mean-square displacement (MSD) $\overline{r^{2}(t)}$ of a point initialized wave packet as a function of time. For the disorder strengths considered, the MSD grows at early times due to transient ballistic or diffusive spreading, and then saturates to a constant well below the system size, indicating the onset of Anderson localization.  This saturation plateau provides a practical criterion for choosing the evolution time. By $t = 10^3$ the MSD has essentially saturated for the system of size $N=512\times512$ with disorder strength $W=14$ and $W=18$ over $7200$ disorder realizations.}
\label{fig:mean_variance_vs_time}
\end{figure}

In this section, we recall the expressions inferred from KPZ physics that govern the spatial structure of long-time wave packets in the localized regime. At sufficiently long times, the disorder-averaged spatial density of the wave packet converges to a stationary, exponentially localized profile. A convenient and robust way to determine the appropriate evolution time for examining the localized wave pakcets is to monitor the disorder-averaged mean-square displacement (MSD) of the wave packet, defined as
\begin{equation} \overline{r^{2}(t)} \equiv 
\overline{\Big\langle \sum_{x,y} (x^{2}+y^{2})\,|\psi^t(x,y)|^{2} \Big\rangle},
\end{equation}
for the point initial condition in Eq~\eqref{Eq:inicirc}. In the localized regime, the MSD eventually saturates to a constant value that remains far smaller than the system size. This saturation plateau provides a clear and robust criterion for selecting the evolution time.  
As shown in Fig.~\ref{fig:mean_variance_vs_time}, once the system is evolved to $t=10^3$, the MSD has almost saturated for the disorder strengths considered, indicating that all transient ballistic or diffusive spreading has ceased. We therefore fix the evolution time to $t=10^3$ for disorder strengths $W>18$, with the understanding that stronger disorder generally requires a shorter evolution time for localizing the wave packets. In the analysis that follows, we omit the explicit time label and use the shorthand notation $|\psi(\boldsymbol r)|^{2} \equiv |\psi^{t}(\boldsymbol r)|^{2}$ for the localized wave packets. 

Our goal is to characterize the fluctuations of the wave-packet tails. Motivated by the analogy with KPZ physics discussed above and in Ref.~\cite{mu_kardar-parisi-zhang_2024} (see also \cite{somoza_universal_2007}), we propose that the logarithmic density takes the form
\begin{equation}
\ln |\psi(\boldsymbol r)|^2 
\underset{t\to\infty,\, r\gg\xi}{\approx}
- \frac{2r}{\xi}
+ \left(\frac{r}{\xi}\right)^\beta \Gamma\, \chi
+ \Lambda,
\label{Eq:stat_dis}
\end{equation}
where $\chi$ is an $O(1)$ random variable, $\Gamma$ and $\Lambda$ are constants, $\xi$ is the localization length, and $\beta = 1/3$ is the KPZ fluctuation growth exponent. The first term describes the standard exponential localization, while the second term encodes the KPZ physics. Crucially, Eq.~\eqref{Eq:stat_dis} is consistent with the single-parameter scaling: the disorder dependence enters only through $\xi$, so plotting the wave-packet profiles in terms of $r/\xi$ should collapse data taken at different disorder strengths onto a universal curve. We verify this property below.

As shown in Ref.~\cite{mu_kardar-parisi-zhang_2024}, Eq.~\eqref{Eq:stat_dis} together with the Tracy-Widom (TW) distribution of fluctuations yield expressions for both the typical and average wave packets. The typical wave packet is
\begin{equation}
\langle \ln |\psi(\boldsymbol r)|^2 \rangle
\underset{r\gg\xi}{\approx}
- \frac{2r}{\xi}
+ \left(\frac{r}{\xi}\right)^{1/3} \Gamma\, \mu
+ \Lambda,
\label{Eq:typical_density}
\end{equation}
where $\mu \approx -1.77$ is the mean of the GUE TW distribution for the point initial condition. Differentiating Eq.~\eqref{Eq:typical_density} gives the stretched-exponential decay,
\begin{equation}
\partial_r \langle \ln |\psi(\boldsymbol r)|^2 \rangle
\underset{r\gg\xi}{\approx}
- \frac{2}{\xi}
+ \frac{\Gamma\,\mu}{3\,\xi^{1/3}}\, r^{-2/3},
\label{Eq:grad_typ_wave}
\end{equation}
whose hallmark is the characteristic $r^{-2/3}$ correction inherited from the KPZ scaling.

The average density $\langle |\psi(\boldsymbol r)|^2 \rangle$, which is more easily accessible experimentally (e.g.\ in cold-atom implementations~\cite{billy_direct_2008, chabe_experimental_2008}), can be obtained by combining Eq.~\eqref{Eq:typical_density} with the TW distribution. Performing this calculation (see Ref.~\cite{mu_kardar-parisi-zhang_2024}) yields
\begin{equation}
\ln \langle |\psi(\boldsymbol r)|^2 \rangle
\underset{r\gg\xi}{\approx}
- \frac{2r}{\xi}
+ \left(\frac{r}{\xi}\right)^{1/3} \Gamma'
+ \left(\frac{r}{\xi}\right)^{2/3} \Gamma''
+ \Lambda',
\label{Eq:ave_density}
\end{equation}
where $\Gamma'$, $\Gamma''$, and $\Lambda'$ are constants determined solely by the typical wave packet and the TW distribution.

\begin{figure}
    \centering
    \includegraphics[width=1\linewidth]{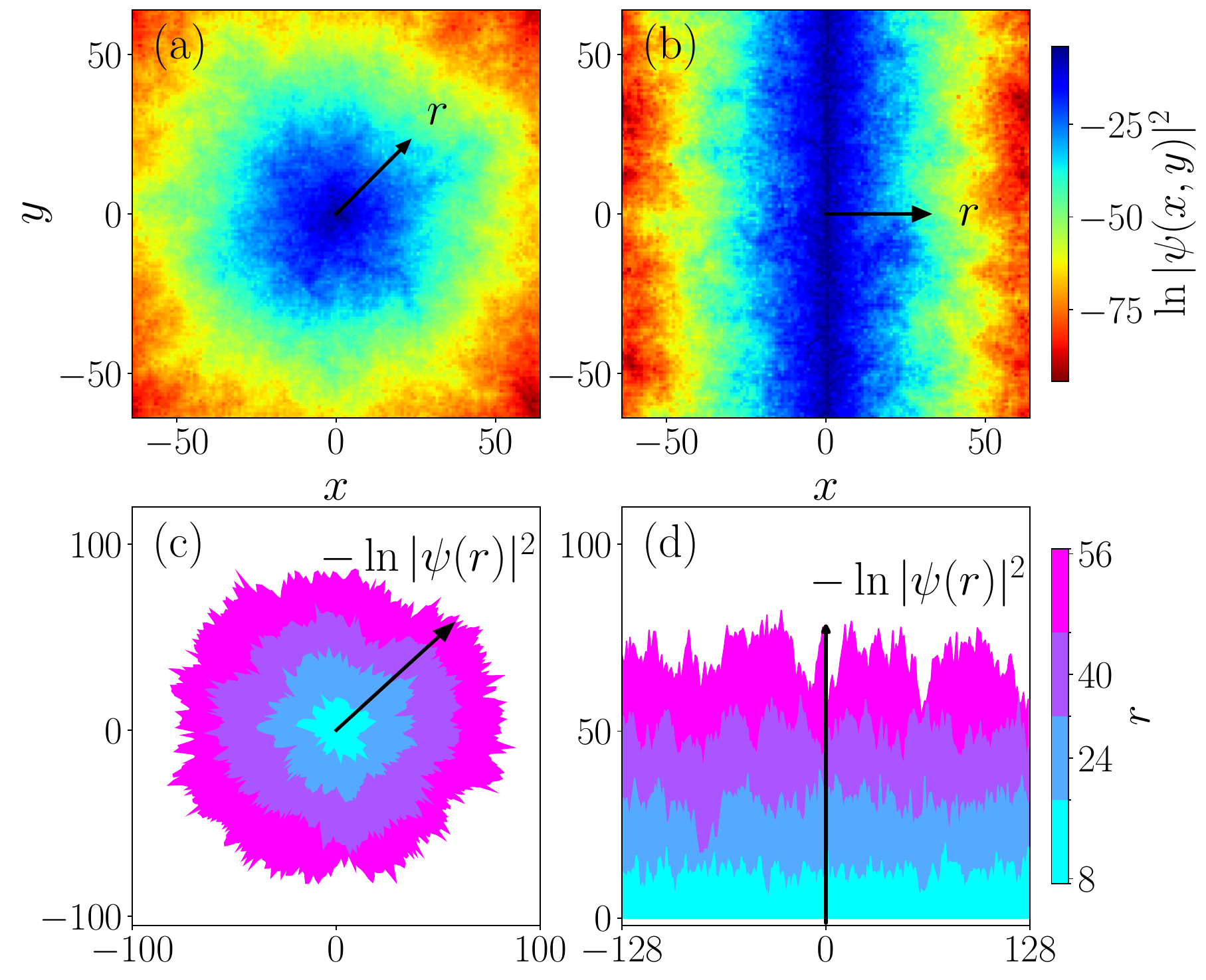}
    \caption{Real-space and mapped representations of the localized wave packets in 2D for the two initial conditions considered (a) point and (b) line. Panels (a) and (b) show $\ln|\psi(x,y)|^{2}$ on the square lattice of size $N=128 \times 128$ with $W=20$ at evolution time $t=10^3$, where arrows indicate the distance $r$ in the localization direction, and color encodes the logarithmic density. Panels (c) and (d) display the corresponding mapped surfaces constructed from the same data: for the point initial condition (c), the magnitude $-\ln|\psi(r)|^{2}$ is plotted in polar form along circles of radius $r$, while for the line initial condition (d) the same quantity is shown along transverse cuts at distances $\pm r$ from the central row. In (c) and (d), color encodes the radius $r$, making the exponential decay of the localized wave function directly visible.
   }
    \label{fig:wavefunction_to_surface}
\end{figure}

\begin{figure*}
\includegraphics[width=0.95\linewidth]{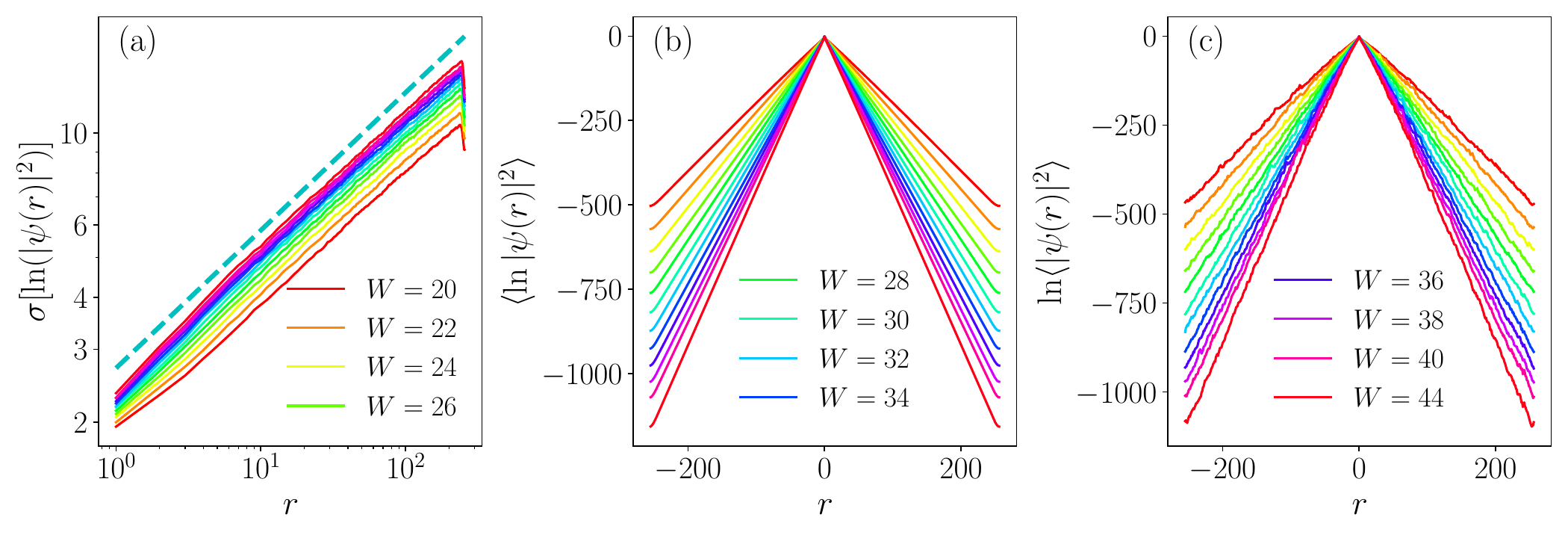}
\caption{Localized wave packets along the diagonal of the square lattice evolved from the point initial condition for various disorder strengths. (a) Standard deviation of $\ln |\psi(\boldsymbol r)|^2$ as a function of distance $r$. The cyan dashed line $\propto r^{1/3}$ is a guide to the eye. 
(b) Typical wave density profiles for different disorder strengths $W$.  
(c) Average wave density profiles for the same disorder strengths.  
Colors indicate the corresponding values of $W$.  
Simulations are performed on a square lattice of size $N=512 \times 512$, evolution time $t=10^3$ and $72000$ disorder realizations.}
\label{fig:localized_wave_W}
\end{figure*}

\subsection{Numerical results for localized wave packets}
\label{numerical_results}

In this section, we present a detailed numerical study of the fluctuations of long-time evolved, localized wave packets in 2D, computed across a broad range of disorder strengths. We begin by examining the spatial profiles of the wave packets and demonstrate how these profiles can be mapped onto a growing rough interface for both the point and line initial conditions, shown in Fig.~\ref{fig:wavefunction_to_surface}. This allows us to apply theoretical insights from KPZ physics. Next we analyze the scaling of the logarithmic density fluctuations in order to validate the expression proposed in Eq.~\eqref{Eq:stat_dis}, with particular emphasis on the fluctuation growth exponent and the stretched-exponential decay. We then extract the localization length from the typical wave packet over a broad range of disorder strengths, enabling a systematic characterization of its disorder dependence. Finally, we test the validity of single-parameter scaling by examining whether the shape of the wave packets, predicted to follow a stretched-exponential form, collapses onto a universal curve when rescaled by the localization length across different disorder strengths. Note that we present only the results for the localized wave packets evolved from the point initial condition, while we have also checked the line initial condition and observed the same behavior.

\subsubsection{Mapping to growing rough interfaces}
We first examine the spatial profiles of the long-time evolved wave packets and establish their mapping to a growing rough interface, providing a basis for identifying universal fluctuations consistent with the KPZ universality class. The mapped interface for the point initial condition is constructed using the same methodology employed for the localized eigenstates. For the line initial condition, we collect the wave density at positions $(x = r, y)$ and plot $-\ln |\psi(x = r, y)|^2$ as a function of $y$ for various values of $r$. The quantity $-\ln |\psi(\boldsymbol r)|^2$ exhibits behavior analogous to the height function of a growing rough surface, as illustrated in Fig.~\ref{fig:wavefunction_to_surface}. In this analogy, the distance $r$ along the localization direction plays the role of time, while the transverse coordinate $y$ corresponds to space in the interface growth process. The wave packet decays exponentially along $r$ while exhibiting fluctuations along $y$. This exponential decay motivates the interpretation of $r$ as time in the surface-growth picture, with the inverse localization length determining the mean growth velocity.

\begin{figure}
\includegraphics[width=0.9\linewidth]{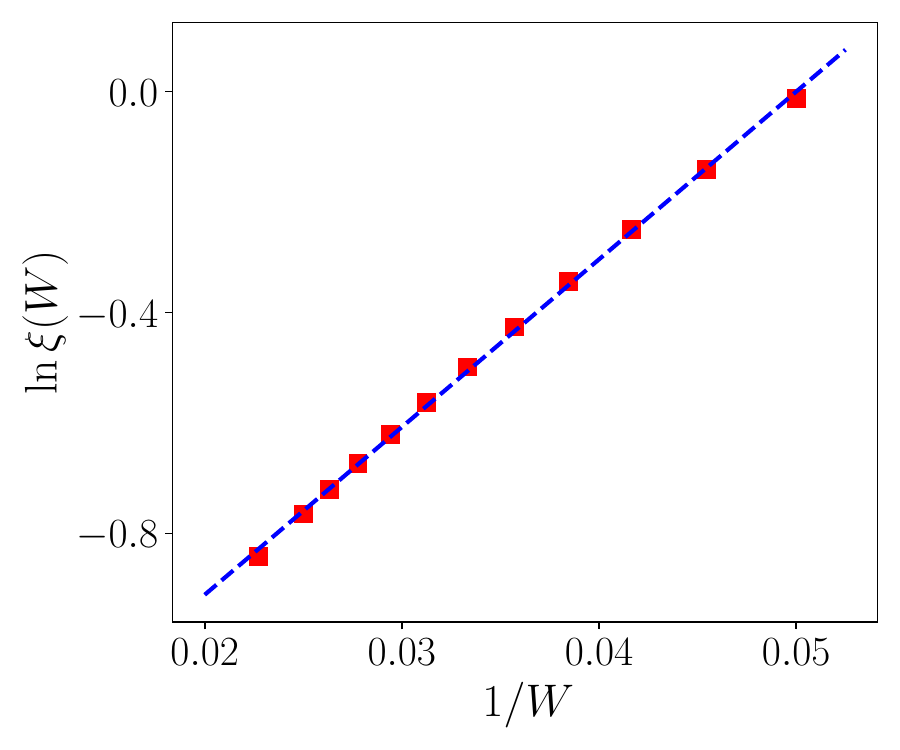}
\caption{Logarithm of the localization length $\ln \xi(W)$ as a function of inverse disorder strength $1/W$, obtained by fitting Eq.~\eqref{Eq:typical_density} with $\Gamma = -1.5$ and $\Lambda = -8.3$ to our numerical results. These constants are determined through a global least-squares fit across all disorder strengths considered. Other parameters are the same as in Fig.~\ref{fig:localized_wave_W}. The blue dashed line corresponds to Eq.~\eqref{eq:loc_vs_w} with $a = 30.4$ and $b = -1.5$ obtained via a linear fit.} 
\label{fig:xi_vs_W}
\end{figure}

\begin{figure}
\includegraphics[width=1.0\linewidth]{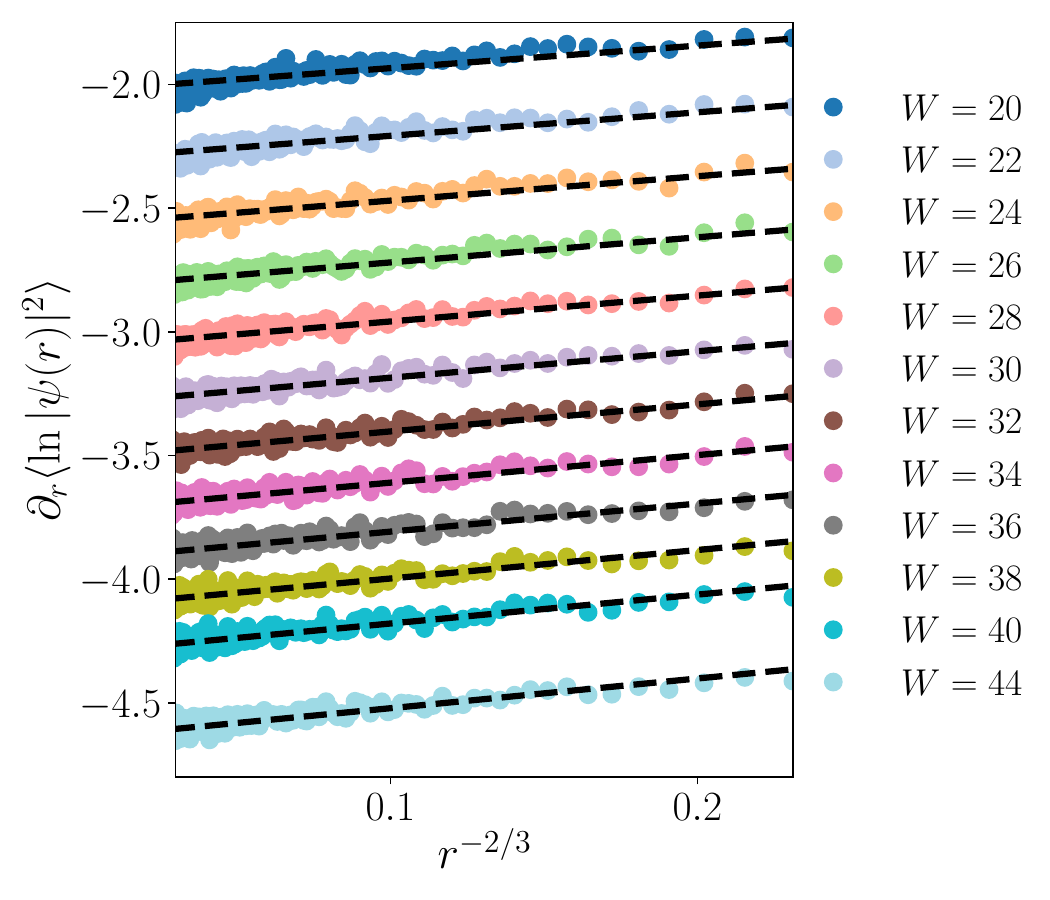}
\caption{Derivative of the typical wave density w.r.t. the distance, $\partial_r\langle\ln|\psi(r)|^2\rangle$, plotted as a function of $r^{-2/3}$. The black dashed line represents Eq.~\eqref{Eq:grad_typ_wave} for different disorder strengths, demonstrating the stretched-exponential behavior of the typical wave density. Other parameters are identical to those of Fig.~\ref{fig:xi_vs_W}.} 
\label{fig:gra_typ_wave}
\end{figure}

\begin{figure*}
\includegraphics[width=0.95\linewidth]{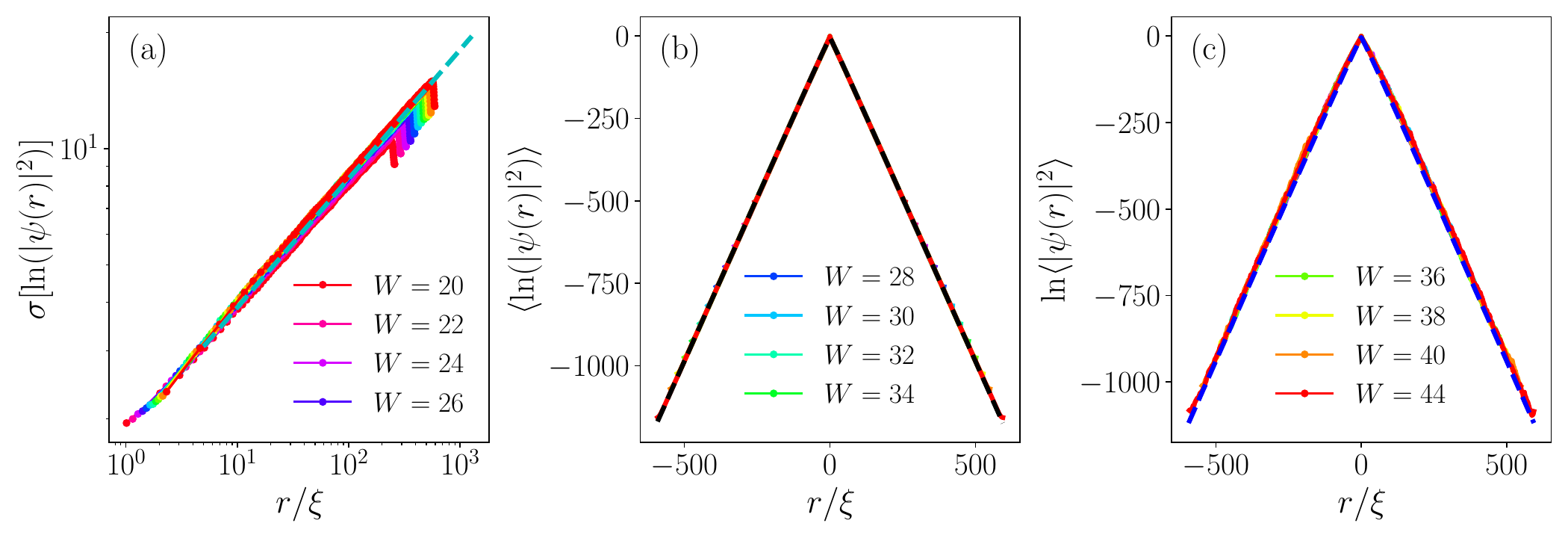}
\caption{Localized wave packets plotted as a function of the rescaled coordinate for various disorder strengths.
(a) Standard deviation of $\ln |\psi(\boldsymbol{r})|^2$ as a function of the rescaled distance $r/\xi$. The cyan dashed line shows the power law $D\,(r/\xi)^{1/3}$ with $D=1.8$.
(b) Typical wave density as a function of $r/\xi$ for different disorder strengths. The black dashed line corresponds to Eq.~\eqref{Eq:typical_density} with $\Gamma = -1.5$ and $\Lambda = -8.3$, obtained from a global least-squares fit to our numerical results across various disorder strengths.
(c) Average wave density as a function of $r/\xi$ for the same set of disorder strengths, with colors indicating the corresponding disorder values $W$. The blue dashed line corresponds to Eq.~\eqref{Eq:ave_density} with $\Gamma'=-0.8$, $\Gamma''=1.1$, and $\Lambda'=-3.0$. Other simulation parameters are identical to those of Fig.~\ref{fig:localized_wave_W}. The excellent agreement between the predictions and the numerical results supports the proposed expressions put forward in Eqs.~\eqref{Eq:typical_density} and~\eqref{Eq:ave_density}.
}
\label{fig:localized_wave_rescaled}
\end{figure*}

\subsubsection{Universal fluctuation scaling in the logarithmic density}

In the same spirit as the study of localized eigenstates, we analyze the scaling of the logarithmic density fluctuations along the localization direction $r$ of the wave packet, which we probe by examining the diagonal of the square lattice.
We show in Fig.~\ref{fig:localized_wave_W}(a) that the standard deviation grows algebraically,
\begin{equation}
\sigma[\ln|\psi(\boldsymbol r)|^2] \sim r^{\beta}.
\label{eq:wave_packet_std}
\end{equation}
We find a fluctuation exponent $\beta \approx 1/3$, in excellent agreement with the KPZ scaling, across a wide range of disorder strengths. This result highlights the emergence of universal KPZ behavior in disordered quantum systems, independent of microscopic details. Note that some deviations appear at large $r$, and we attribute them to the boundary effect. 

\subsubsection{Typical and average wave densities and the single-parameter scaling}

We now turn to the spatial profiles of the typical and average wave densities for various disorder strengths, shown in Fig.~\ref{fig:localized_wave_W}(b) and (c). We aim to test the single-parameter scaling hypothesis \cite{abrahams_scaling_1979, evers_anderson_2008, pichard1981finite, mackinnon1981one} for both the typical and average wave densities. Using the expressions in Eqs.~\eqref{Eq:typical_density}, we determine the constants $\Gamma$, $\Lambda$, and the localization length $\xi$ via a global least-squares fitting procedure across all the disordered strengths considered. In 2D, the localization length is expected to follow the exponential dependence on the inverse disorder strength
\begin{equation}
\ln \xi = \frac{a}{W} + b,
\label{eq:loc_vs_w}
\end{equation}
where $a$ is a constant and $b$ accounts for subleading corrections \cite{mackinnon1981one, manai2015experimental, vsuntajs2023localization}. Our numerical results for $\xi$ across disorder strengths $W$, shown in Fig.~\ref{fig:xi_vs_W}, display excellent agreement with Eq.~\eqref{eq:loc_vs_w}. In particular, we demonstrate the nontrivial stretched-exponential behavior predicted by Eq.~\eqref{Eq:grad_typ_wave} in Fig.~\ref{fig:gra_typ_wave}, thereby confirming the expression in Eq.~\eqref{Eq:typical_density}.

Finally, we perform the central test of the single-parameter scaling hypothesis: whether both the typical and average wave densities collapse onto universal curves when expressed as functions of the rescaled distance $r/\xi$. We first examine the fluctuation scaling and observe excellent collapse in Fig.~\ref{fig:localized_wave_rescaled}(a). Remarkably, the typical and average wave densities also collapse very well across all disorder strengths considered, as shown in Fig.~\ref{fig:localized_wave_rescaled}(b) and (c). This robust collapse provides compelling evidence for the single-parameter scaling. It also validates the proposed expressions in Eq.~\eqref{Eq:typical_density} and Eq.~\eqref{Eq:ave_density}, showing that these predictions motivated by the KPZ physics accurately capture both the fluctuations and the shape of localized wave packets in the strong-disorder regime of 2D Anderson localization.

\section{Conclusions}
\label{conclusion}

In this work we investigated two complementary manifestations of strong Anderson localization in two dimensions: the properties of individual eigenstates and localized wave packets. By combining extensive exact-diagonalization studies with large-scale unitary dynamics, significantly extending previous investigations \cite{nguyen_aaronov-bohm_1985,medina_quantum_1992,pietracaprina_forward_2016,prior_conductance_2005,somoza_universal_2007,lemarie_glassy_2019,mu_kardar-parisi-zhang_2024,Mu_prb_2025,swain_2d_2025}, we demonstrated that both static and dynamical observables in 2D Anderson localization exhibit universal fluctuations governed by the KPZ universality class \cite{kardar_dynamic_1986, takeuchi_appetizer_2018}, as well as glassy features characteristic of pinned elastic manifolds \cite{mezard_glassy_1990, fisher1991directed, wiese2022theory}. This provides a unified physical picture for fluctuations and microscopic structure in Anderson localization beyond the exactly solved one-dimensional case.

Our first main result concerns the spatial fluctuations and microscopic structure of individual eigenstates. We showed that the logarithmic density of a localized eigenstate obeys the KPZ scaling~\cite{kardar_dynamic_1986, takeuchi_appetizer_2018}, and can thus be naturally interpreted as a KPZ height field growing with the distance from the localization center. Pushing this correspondence toward directed-polymer physics, we resolved the internal geometric structure of eigenstates at the microscopic level. The wavefunction amplitude at a distant site is dominated by a single optimal path. We demonstrated that this dominant path exhibits the glassy behavior familiar from pinned elastic manifolds~\cite{mezard_glassy_1990, fisher1991directed, wiese2022theory}: its transverse coordinate forms plateaus separated by jumps as the observation point is varied, reflecting pinning to favorable spatial regions punctuated by abrupt rearrangements akin to avalanche. Importantly, these jumps correlate directly with the inhomogeneous localization profile of the eigenstate. The eigenstate typically organizes into several “branches” where localization is weaker, separated by regions of stronger localization. The pinned configurations of the dominant path coincide with these branches, while avalanche-like rearrangements occur in the strongly localized regions that separate one branch from the next.

We then examined dynamical observables by analyzing wave packets evolved to long times. The stretched-exponential envelope predicted from KPZ physics, previously demonstrated numerically for the quantum kicked rotor \cite{mu_kardar-parisi-zhang_2024}, was tested across a broad range of disorder strengths wherever the localization tails are accessible with the state-of-the-art numerical routine. We showed unambiguously that single-parameter scaling holds \cite{abrahams_scaling_1979, evers_anderson_2008, pichard1981finite, mackinnon1981one}.

Taken together, our results suggest a unified emerging description: the logarithmic density of localized states in 2D Anderson localization behaves as a rough interface growing with distance, while the detailed spatial structure of the localized wavefunction is governed by an effective directed-polymer problem exhibiting characteristic glassy features. This conceptual bridge between disordered quantum systems and classical stochastic-growth processes provides a new source of intuition as well as quantitative predictive power.

Looking forward, several directions appear particularly promising. Rare events and extremal fluctuations, which are central to the KPZ and directed polymer problems \cite{derrida1988polymers, hartmann2018high, monthus2006probing}, warrant a more systematic quantification in the context of Anderson localization. Approaches rooted in extreme-value statistics theory could sharpen our understanding of these atypical but decisive contributions, especially in regimes dominated by strong disorder, see \cite{swain_2d_2025, PhysRevB.105.094202, miranda2025large} for recent studies in 2D and on random graphs. A second challenge concerns fluctuations in three or higher-dimensional Anderson localization. The $(d+1)$-dimensional (complex) directed polymer problem is inherently richer than the $(1+1)$-dimensional case, see e.g. \cite{monthus2006statistics, PhysRevB.110.L060202} and refs.~therein. This analogy with KPZ and directed polymer physics could help clarifying the role of dimensionality in the Anderson transition at strong disorder, an important open question actively debated \cite{arenz2023wegner, zirnbauer2023wegner, PhysRevB.109.174216, PhysRevB.109.L220202, tonetti2025testing}. Finally, extending this perspective to interacting or driven systems, including many-body localized phases and periodically driven (Floquet) settings, may uncover new organizing principles where directed-polymer methods and KPZ physics provide unifying intuition \cite{PhysRevB.105.094202, miranda2025large}.

\section{Acknowledgments}
We acknowledge the use of computational resources at the Singapore National Super Computing Centre (NSCC) ASPIRE-2A cluster, the Calcul en Midi-Pyrénées (CALMIP) and Max Planck Institute for the Physics of Complex Systems (MPI-PKS) for our numerical simulations. 
This work is supported by the Singapore Ministry of Education Academic Research Funds Tier II (MOE-T2EP50223-0009 and MOE-
T2EP50222-0005), the ANR projects QUTISYM
(ANR-23-PETQ-0002) and ManyBodyNet, the EUR Grant NanoX No. ANR-17-EURE-0009. GL acknowledges InPhyNi (CNRS and UCA) for hospitality.
J.G. acknowledges support by the National Research Foundation, Singapore through the National Quantum Office, hosted in A*STAR, under its Centre for Quantum Technologies Funding Initiative (S24Q2d0009).

\bibliography{references.bib}

\end{document}